\documentclass[prx,preprint,longbibliography,superscriptaddress,aps]{revtex4-1}
\usepackage{graphicx}
\usepackage{array}
\usepackage{amssymb}
\usepackage{amsfonts}
\usepackage{amsmath}
\usepackage{mathrsfs}
\usepackage{color}
\usepackage{booktabs}
\usepackage{threeparttable}
\usepackage{multirow}
\usepackage{subfigure}
\usepackage{epsfig}
\usepackage{threeparttable}
\usepackage{chngpage}
\usepackage{float}
\usepackage{color}
\usepackage{bm}
\usepackage{dcolumn}
\usepackage{textcomp}
\usepackage{verbatim}
\usepackage{epstopdf}
\usepackage{multirow}
\usepackage{tabularx}
\usepackage[labelfont=bf]{caption}
\DeclareCaptionLabelSeparator{vline}{ $|$ }
\usepackage{caption}
\makeatletter
\renewcommand\@biblabel[1]{#1.}
\makeatother

\begin{document}

\title{The paradox of controlling complex networks: control inputs versus
energy requirement}

\author{Yu-Zhong Chen}
\affiliation{School of Electrical, Computer, and Energy Engineering,
Arizona State University, Tempe, AZ 85287, USA}

\author{Lezhi Wang}
\affiliation{School of Electrical, Computer, and Energy Engineering,
Arizona State University, Tempe, AZ 85287, USA}

\author{Wenxu Wang}
\affiliation{Department of Systems Science, Beijing Normal University,
Beijing, 10085, China}

\author{Ying-Cheng Lai} \email{Ying-Cheng.Lai@asu.edu}
\affiliation{School of Electrical, Computer, and Energy Engineering,
Arizona State University, Tempe, AZ 85287, USA}
\affiliation{Department of Physics, Arizona State University, Tempe,
AZ 85287, USA}


\begin{abstract}

One of the most challenging problems in complex dynamical systems
is to control complex networks. In previous frameworks based on the
structural or the exact controllability theories, the ability to
steer a complex network toward any desired state is measured by the
minimum number of required driver nodes. However, if we implement
actual control by imposing input signals on the minimum set of
driver nodes as determined, e.g., by the structural controllability
theory, an unexpected phenomenon arises: the energy required to
approach a target state with reasonable precision is often unbearably 
large, precluding us from achieving actual control, i.e., the designated 
state can not be reached in effect, especially for networks with a 
small number of drivers. In particular, the energy of controlling a set
of networks with similar structural properties follows a fat-tail
distribution, indicating the existence of networks with practically
divergent energy. We aim to reconcile the paradox of controlling
complex networks: optimal structural controllability versus
unrealistic energy required for control. We identify fundamental
structural ``short boards'' in complex networks that play a
dominant role in the enormous energy, and offer a theoretical
interpretation for the fat-tail energy distribution and simple
strategies to significantly reduce the energy by imposing slightly
augmented set of input signals on properly chosen nodes. Our
findings indicate that, although full control can be guaranteed by
the prevailing structural controllability theory, it is necessary
to balance the number of driver nodes and the control energy to
achieve actual control, and our results provide a framework to
address this outstanding issue.

{\bf Notes on the submission history of this work}: This work started
in late 2012. The phenomena of power-law energy scaling and energy
divergence with a single controller were discovered in 2013.
Strategies to reduce and optimize control energy was articulated
and tested in Spring 2014. The senior co-author (YCL)
gave talks about these results at several conferences, including
the NETSCI 2014 Satellite entitled ``Controlling Complex Networks''
on June 2. The paper was submitted to PNAS in September 2014 and
was turned down. It was revised and submitted to PRX in early
2015 and was rejected. After that it was revised and submitted
to Nature Communications in May 2015 and again was turned down.

\end{abstract}

\maketitle

\section{Introduction} \label{sec:intro}

The past fifteen years have witnessed tremendous advances in our understanding
of complex networked structures in various natural, social, and technological systems,
as well as the dynamical processes taking place on them\cite{WS:1998,BA:1999,AJB:1999,
ASBS:2000,AJB:2000,CEBH:2000,JMBO:2001,PV:2001,NWS:2002,AB:2002,Newman:2003,
PDFV:2005,BLMCH:2006,Caldarelli:2007,NABV:2010,Fortunato:2010}.
The significant issue of control arises naturally, but this remains to be
outstanding and extremely challenging, since nonlinear dynamical processes
generally take place on complex networks. Control of nonlinear dynamics,
especially when chaos is present, can be done but only for low-dimensional
systems~\cite{OGY:1990,BGLMM:2000}. Despite the development of nonlinear control
methods~\cite{SL:book,WC:2002,WS:2005,SBGC:2007,YCL:2009,RME:2009} in
certain particular situations such as consensus~\cite{EMCCB:2012},
communication~\cite{KMT:1998,CLCD:2007}, traffic~\cite{Srikant:book} and
device networks~\cite{Luenberger:book,SL:book}, a general framework of
controlling complex nonlinear-dynamical networks has yet to be developed.
A natural approach is to reduce the problem to controlling complex networks
with linear dynamics based on traditional frameworks from control
engineering~\cite{Kalman:1963,Lin:1974,SP:1976,RW:1997,Sontag:book}.

In the past a few years, great progress was made toward understanding
the linear controllability of complex networks in terms of the fundamental
issue of the minimum number of driver nodes required to steer the whole
network system from an arbitrarily initial state to an arbitrarily final state
in finite time~\cite{LH:2007,LSB:2011,WNLG:2012,NV:2012,YRLLL:2012,LSB:2012,NA:2012}.
In particular, Liu et al. successfully adopted the classic structural
controllability theory developed by Lin~\cite{Lin:1974} to complex networks
of various topologies~\cite{LSB:2011}, for which the traditional
Kalman's rank condition~\cite{Kalman:1963} is difficult to be
applied~\cite{LH:2007}. The ground breaking results show that, the structural
controllability of a directed network can be assessed by using the
maximum matching~\cite{HK:1973,ZOY:2003,ZM:2006} algorithm.
The effects of the density of in/out degree nodes were incorporated into
the structural controllability framework~\cite{MDB:2014}, which
has also been applied to protein interaction networks~\cite{Wuchty:2014}.
Recently, based on the classic Popov-Belevitch-Hautus (PBH) rank
condition~\cite{PBH} in traditional control engineering, a variant of the
structural-controllability theory, the so-called exact controllability
framework, was developed~\cite{YZDWL:2013}.

For both the structural- and exact-controllability frameworks, the
aim is to determine the minimum number of driver nodes, $N_\text{D}$, for
networks of various topologies. However, we have encountered unexpected
difficulties in carrying out {\em actual} control of complex networks
by using the minimum set of driver nodes as determined by the 
controllability frameworks. This concerns effectively the issue of guiding
the network system to approach a final state with acceptable proximity error. 
In particular, given an arbitrary complex network, once $N_\text{D}$ is 
determined, we can calculate the specific control signals by using the 
standard linear systems theory~\cite{Rugh:book} and apply them at various
unmatched nodes. A surprising finding is that, quite often, the actual
control of the system is difficult to be achieved computationally in
the sense that in any finite time, it is not possible to drive the system
from an arbitrary initial state to an arbitrary final state, i.e., the 
actual state the system finally reaches is unreasonably far from the 
designated one. This difficulty in realizing actual control, which has 
not been formerly addressed in any other works, persists for a large 
number of model and real-world networks, prompting us to study if the 
developed controllability frameworks can ensure actual control with given 
finite computational precision and, more importantly, to consider the 
issue of control energy.

In this paper, we investigate the issue of control energy in the framework
of structural controllability theory. We find that, the energy required to
steer a system from a specific initial state to a target state in finite
time follows a fat-tail distribution, indicating the existence of
extraordinarily high energy requirement. In extreme but not uncommon cases, 
the energy is practically divergent. This phenomenon signifies the emergence 
of a {\em paradox} in controlling complex networks: although a small 
fraction of driver nodes can guarantee full control of the network system 
mathematically, the energy required to achieve control is often unbearable. 
We resolve the paradox by presenting the idea of control chains, in which 
the fat-tail distribution of the energy can be derived as a key structural 
feature. The theory of control chains enables us to offer simple strategies 
to significantly reduce the control energy through small augmentation of
the number of control signals beyond $N_\text{D}$. In this regard, the 
quantity $N_\text{D}$, on which the structural controllability theories 
focus, can effectively be regarded as the lower bound of the actual number 
of control signals required. To realize actual control of a complex network,
it is imperative to find the trade-off between the number of external input
signals and feasible energy consumption.

{\em Remark.} In Ref.~\cite{YRLLL:2012}, partial theoretical bounds for 
the control energy were derived. The bounds are partial because, for 
example, for networks whose lower bounds can be obtained, the upper 
bounds typically diverge. This property of divergence was puzzling:
does it mean that the actual energy required would diverge as well and, 
if so, can a complex network actually be controlled? The present work
was largely motivated by these questions, in which we obtain a detailed 
understanding of the physically important issue of practical 
controllability of complex networks through the discovery of a general 
scaling law for the distribution of the energy required for control. The
existence of control chain is also uncovered, enabling us to articulate 
practical strategies to significantly reduce the control energy. 

\section{Control Formulation and Implementation}

\paragraph*{Optimal control energy and Gramian matrix.}
To calculate the optimal energy required to control a complex network
in the framework of structural controllability, we consider the standard
setting of linear dynamical systems under control
input~\cite{LH:2007,LSB:2011,YZDWL:2013}:
\begin{equation} \label{eq:xt}
\mathbf{\dot{x}}=A\mathbf{x} + B\mathbf{u},
\end{equation}
where $\mathbf{x}=[x_1(t),\ldots,x_N(t)]^T$ is the state variable of the
whole network system, the vector $\mathbf{u}=[u_1(t),\ldots,u_M(t)]^T$ is
the control input or the set of control signals, $A=\{a_{ij}\}$ is the $N\times N$
adjacency matrix of the network, and $B=\{b_{ik}\}$ is the $N\times N_\text{D}$
control matrix specifying the set of ``driver'' nodes~\cite{LSB:2011},
each receiving a control signal (corresponding to one component of the control
vector $\mathbf{u}$). The minimum number $N_\text{D}$ of driver nodes to fully
control a network is determined through the set of maximum matching paths~\cite{LSB:2011}.
A node is chosen to be the driver node if it is the starting point of a maximum
matching path. The system is fully controlled only if each and every node is
either a driver node or being driven along a maximum matching path.
Optimal control of a linear network in the sense that the energy is minimized
can be achieved when the input control signals $\mathbf{u}_t$
are chosen as~\cite{Chen:book}:
\begin{equation} \label{eq:ut}
\mathbf{u}_t = B^T \cdot e^{A^T (t_\text{f}-t)} \cdot
W^{-1} \cdot (\mathbf{x}_{t_\text{f}}-e^{A{t_\text{f}}}
\cdot \mathbf{x}_0),
\end{equation}
where
\begin{equation} \label{eq:Gramian}
W \equiv \int_{t_0}^{t_\text{f}} e^{A\tau}
B \cdot {B^T} \cdot e^{A^T\tau}d\tau
\end{equation}
is the Gramian matrix, a positive-definite and symmetric matrix~\cite{Rugh:book},
which is the base to determine, quantitatively, if a system is actually
controllable. In particular, the system is controllable only when $W$ is
nonsingular (invertible)~\cite{Rugh:book,Chen:book}.

Given matrices $A$ and $B$, the initial and the final (target) states of the
system as well as the control time $t_\text{f}$, the control vector $\mathbf{u}$
can be determined in a standard manner~\cite{Rugh:book} via the Gramian matrix
$W$. The energy required through the control input $\mathbf{u}$ is given
by~\cite{Rugh:book}
\begin{equation} \label{eq:energy}
E(t_\text{f})=\int_{0}^{t_\text{f}}\mathbf{u}^T_{t}\mathbf{u}_tdt,
\end{equation}
where control is initiated at $t = 0$.

\paragraph*{Numerical implementation of control.}
We use the Erdos-Renyi (ER) type of directed random networks~\cite{ER:1959,ER:1960}
and the Barab\'asi-Albert (BA) type of directed scale-free networks~\cite{AB:2002}
with a single parameter $P_\text{b}$. Specifically, for a pair of nodes $i$ and $j$
with a link, the probability that it points from the smaller-degree to the
larger-degree nodes is $P_\text{b}$, and $1-P_\text{b}$ is the probability that the
link points in the opposite direction (if both nodes have the same degree, the
link direction is chosen randomly). (See Appendix A for
analytical treatment of the in- and out-degree distributions.) To determine
the set of driver nodes, we use the maximum-matching algorithm~\cite{Lin:1974},
which gives the control matrix $B$. For each combination of $A$ and $B$, we first
randomly choose the initial and final states. We then calculate the corresponding
Gramian matrix $W$ [Eq.~(\ref{eq:Gramian})], the input signal $\mathbf{u}_t$
[Eq.~(\ref{eq:ut})], the actual final states $\mathbf{x}_{t_\text{f}}^{\star}$
[Eq.~(\ref{eq:xt})], and finally the control energy $E(t_\text{f})$
[Eq.~(\ref{eq:energy})]. Repeating this process for each and every independent
network realization in the ensemble entails an extensive statistical analysis
of the control process.

\section{Resolution of control paradox and control-energy distribution} 
\label{sec:paradox}

\subsection{Quantification of control energy and resolution of control paradox}

Mathematically, if the Gramian matrix $W$ is singular, the energy diverges.
Through extensive and systematic numerical computations, we find that, even
when $W$ is non-singular in the mathematical sense, for typical complex networks
its condition number can be enormously large~\cite{SM:2013}, making it
effectively singular as any physical measurement or actual computation must
be associated with a finite precision. Say in a physical experiment the
precision of measurement is $\varepsilon$. In a computational implementation
of control, $\varepsilon$ can be regarded as the computer round-off error.
Consider the solution vector $\mathbf{X}$ of the linear equation:
$W\cdot \mathbf{X} = \mathbf{Y}$, where $\mathbf{Y}$ is a known vector. Let
$C_W$ be the condition number of $W$. The accuracy of the numerical solution
of $\mathbf{X}$, denoted by $e_X = 10^{-k}$ ($k$ is a positive integer), is
bounded by the product between $C_W$ and $\varepsilon$~\cite{Strang:book}.
We see that, if $C_W$ is larger than $10^{-k}/\varepsilon \equiv \bar{C}_W$,
it is not possible to bring the system to within $10^{-k}$ of the final state,
so control cannot be achieved in finite time.

For a large number of networks drawn from an ensemble of networks with a
pre-defined topology, the condition numbers of their Gramian matrices are
often orders of magnitude larger than $\bar{C}_W$ (see Fig.~A1 in 
Appendix B1 for the relation between $C_W$ and $e_X$). For these networks,
not only is the control vector unable to drive the system to the target
state, but the associated energy can be extremely large. These observations
suggest the following criterion to define controllability in terms of the
control energy: a network is controllable with respect to a specific control
setting if and only if the condition number of its Gramian matrix is less
than $\bar{C}_W$, a critical number determined by both the measurement or
computational error and the required precision of control. Quantitatively,
for a given set of network parameters (hence a given network ensemble) and
control setting, the probability that the condition number of the Gramian
matrix is less than $\bar{C}_W$, $P(\bar{C}_W)$, can effectively serve as a
new type of controllability, which we name as {\em practical controllability}.
Increasing the precision of the computation, e.g., by using special simulation
packages with round-off error orders of magnitude smaller than that associated
with the conventional double-precision computation, would convert a few
uncontrollable cases into controllable ones, but vast majority of the
uncontrollable cases remain unchanged.

\begin{figure}
\centerline{\includegraphics[width=\linewidth]{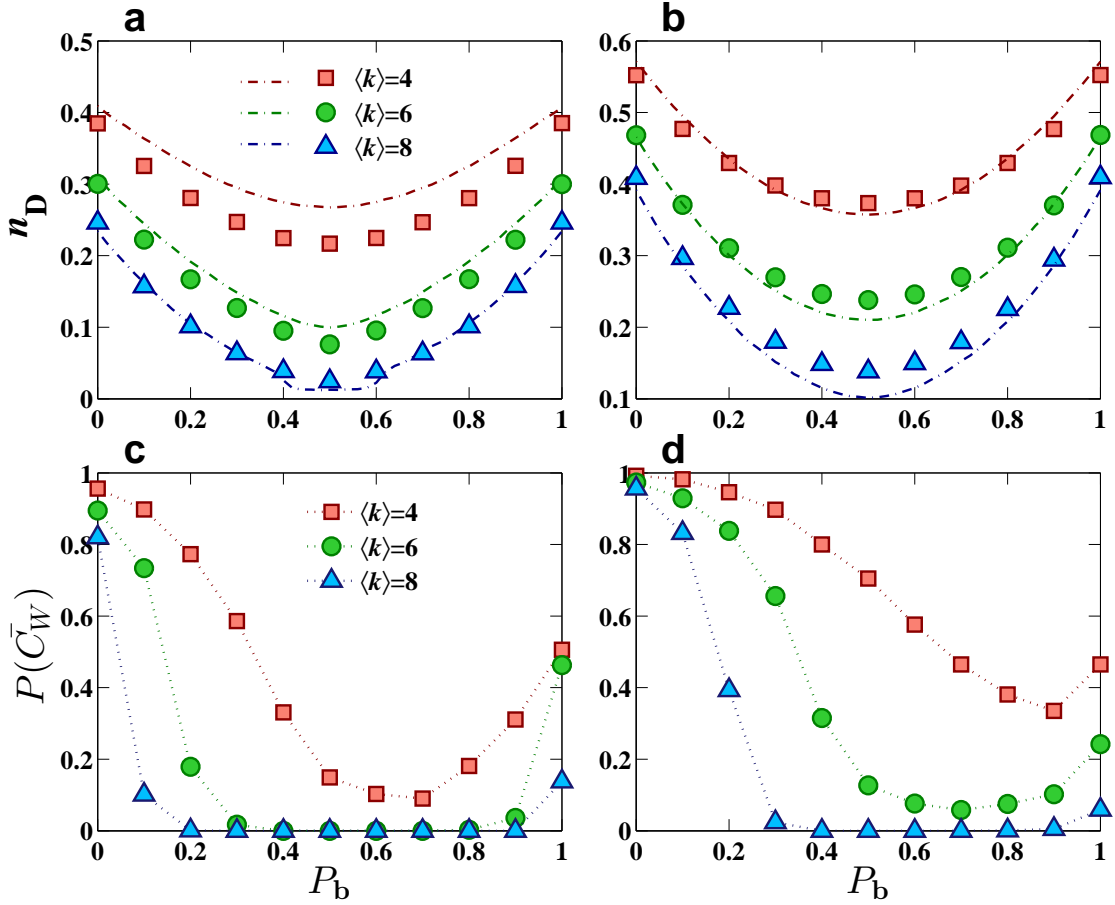}}
\caption{\textbf{Structural and practical controllability measures in
directed networks}. Structural controllability measure $n_{\text{D}}$
versus directional edge probability $P_\text{b}$ for (a) ER random
networks and (b) BA scale-free networks of size $N = 1000$ and
three values of the average degree ($\langle k\rangle=4$, $6$, and $8$).
The dash-dotted lines represent the results obtained by the cavity
method~\cite{LSB:2011}, and the squares, triangles, and circles
are the simulation results from the maximum matching
algorithm~\cite{LSB:2011}. (c,d) Measure of practical controllability
$P(\bar{C}_W)$ for ER random and BA scale-free networks of size
$N=100$, respectively, where $P(\bar{C}_W)$ is the probability that
the condition number of the Gramian matrix is less than some
physically reasonable threshold value versus $P_\text{b}$. Comparing
(a) with (c), or (b) with (d), we observe the striking phenomenon that,
in the parameter regime where the number of driver nodes is minimized
so that the corresponding networks are deemed to be most structurally
controllable, they are practically uncontrollable. The phenomenon
persists regardless of the network size and type.}
\label{fig:main}
\end{figure}

Figures~\ref{fig:main}(a-b) show the percentage of driver nodes
$n_{\text{D}} \equiv N_\text{D}/N$ versus the directional link probability
$P_\text{b}$. We see that $n_{\text{D}}$ is minimized for $P_\text{b} \approx 0.5$,
indicating that, mathematically, only a few control signals are needed to control
the whole network, leading to optimal structural controllability. But can
practical controllability be achieved in the same parameter regime where
the structural controllability is optimized?

Figure~\ref{fig:main}(c) show, for the same networks as in Fig.~\ref{fig:main}(a),
the measure of control energy, i.e., the probability $P(\bar{C}_W)$, versus the
network parameter $P_\text{b}$. We see that, for both regimes of small and large
$P_\text{b}$ values where the structural controllability is weak [corresponding
to relatively high values of $n_{\text{D}}$ in Fig.~\ref{fig:main}(a)], the
practical controllability is relatively strong. In the regime of small $P_\text{b}$
values, most directed links in the network point from small- to
large-degree nodes. In this case, the network is more practically controllable,
in agreement with intuition. The surprising result is that, in the regime of
intermediate $P_\text{b}$ values (e.g., $P_\text{b}$ around 0.5) where the number
of driver nodes to control the whole network is minimized so that the structural
controllability is regarded as strong, the practical controllability is in fact
quite weak, as the probability of the condition number being small is close to
zero. For example, for $\langle k\rangle=4$, the minimum value of $P(\bar{C}_W)$
is only about 0.1 for $P_\text{b} \approx 0.6$; for $\langle k \rangle=6$ and
$\langle k \rangle=8$, the minimum values are essentially zero. A striking
phenomenon is that the minimum value of $P(\bar{C}_W)$ occurs in a wide range of
the parameter $P_\text{b}$, e.g., $[0.3,0.8]$ and $[0.2,0.9]$ for
$\langle k \rangle=6$ and $\langle k \rangle=8$, respectively. This indicates
that the network is practically uncontrollable for most cases where the
structurally controllability is deemed to be optimal. The same phenomenon
holds for different network sizes (see Fig.~A2 in Appendix B2).
Another interesting finding in Fig.~\ref{fig:main} is that $N_\text{D}$
is symmetric about $P_\text{b}=0.5$. However, the symmetry is broken for
$P(\bar{C}_\text{W})$, indicating that there is no simple negative correlation
between $N_\text{D}$ and $P(\bar{C}_\text{W})$. This prompts us to find more
essential structural properties responsible for the smallness of
$P(\bar{C}_\text{W})$.

\begin{figure}
\centerline{\includegraphics[width=\linewidth]{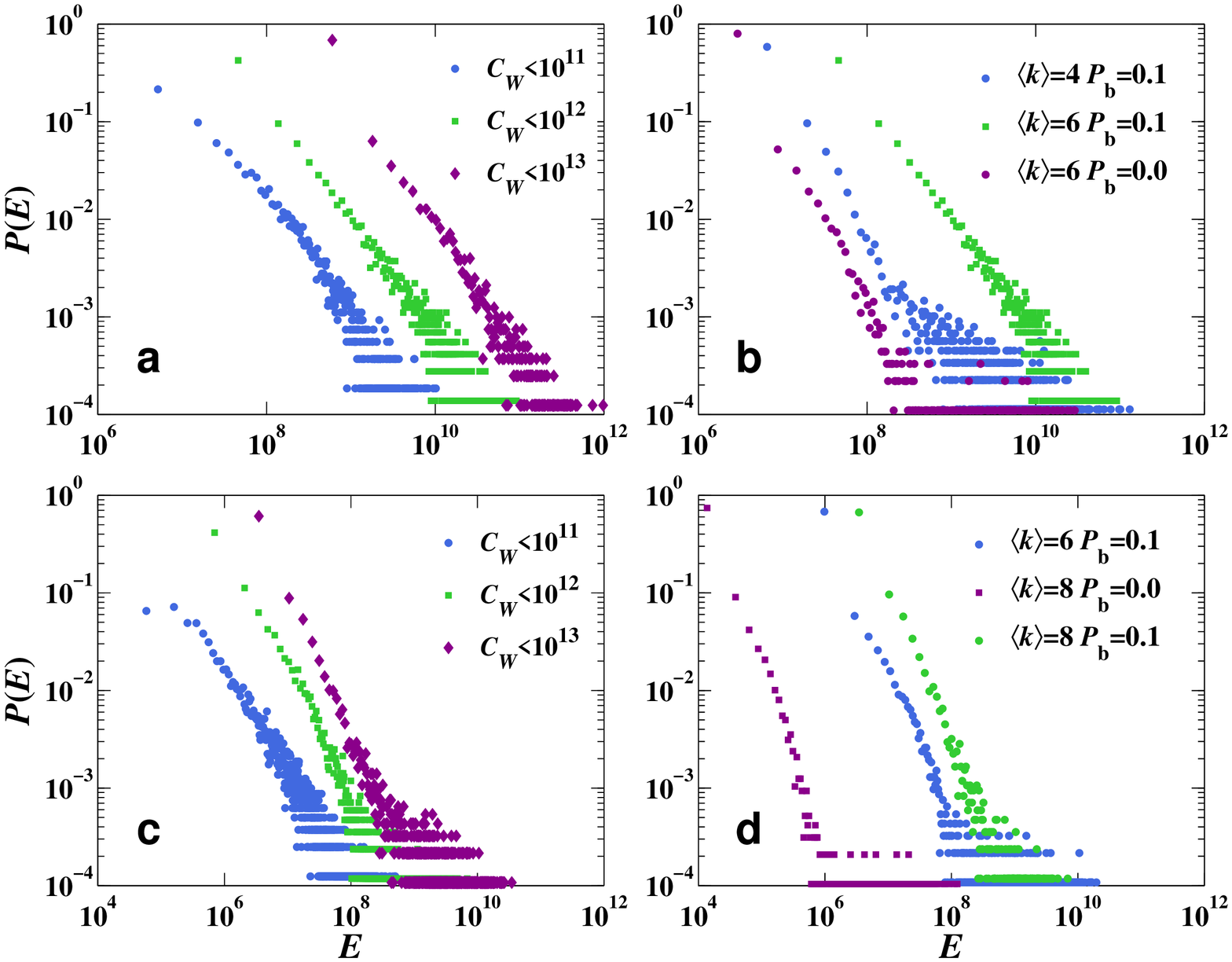}}
\caption{\textbf{Distributions of control energy for practically controllable
networks}.
(a,c) Energy distributions under different values of the threshold
condition number $\bar{C}_W$ for $P_{\text{b}}=0.1$ and $t_{\text{f}}=1$.
$\langle k\rangle=6$ for random networks (a) and $\langle k\rangle=8$ for scale-free networks (c).
(b,c) Energy distributions for different values of the average degree
$\langle k\rangle$ and the directional connection
probability $P_\text{b}$ under $\bar{C}_W = 10^{12}$ and $t_{\text{f}}=1$
for ER random and BA scale-free networks, respectively.
In all cases, we observe an algebraic (power-law) scaling behavior.}
\label{fig:energy_distri}
\end{figure}

\subsection{Concept of control chains and distribution of control energy}

Suppose the network is practically controllable so that the required control
energy is not unrealistically large. For an ensemble of randomly realized network
configurations with the same structural properties and for different control
settings, the control energy can be regarded as a random variable. What is then
its probability distribution? To gain insights, we generate directed networks
with different values of $\langle k\rangle$ and $P_\text{b}$. We then implement
the maximum matching algorithm~\cite{LSB:2011} to obtain the control matrix $B$ and
calculate the minimum energy by using Eq.~(\ref{eq:energy}) for final time $t_\text{f}$.
For each network, the initial states $\mathbf{x_0}$ and desired final states
$\mathbf{x}_{t_\text{f}}$ are randomly chosen. The calculation of energy is done
only for those networks with condition number smaller than $\bar{C}_W$, and a variety of $\bar{C}_W$ values are adopted.
Representative results are shown in Fig.~\ref{fig:energy_distri}, where an
algebraic (power-law) scaling behavior with fat tails is observed for all cases
with the scaling exponent approximately equal to $1.5$. The scaling is robust
against various $\bar{C}_W$ values [Figs.~\ref{fig:energy_distri}(a) and (c)]
and network sizes (see Fig.~A3 in Appendix B3).
From Figs.~\ref{fig:energy_distri}(a) and (c), we see that
different values of $\bar{C}_W$ result in different groups of
practically controllable networks, and the required control energy
in general increases with $\bar{C}_W$.
In Figs.~\ref{fig:energy_distri}(b) and (d), the value of
$\bar{C}_W$ is fixed and the control energy required is larger for larger
value of $P_\text{b}$ as compared with the case of $P_\text{b} = 0$. This is
intuitively correct as, for $P_\text{b} = 0$, all directed links point from
small- to large-degree nodes, facilitating control of the whole network.

\begin{figure}
\centerline{\includegraphics[width=\linewidth]{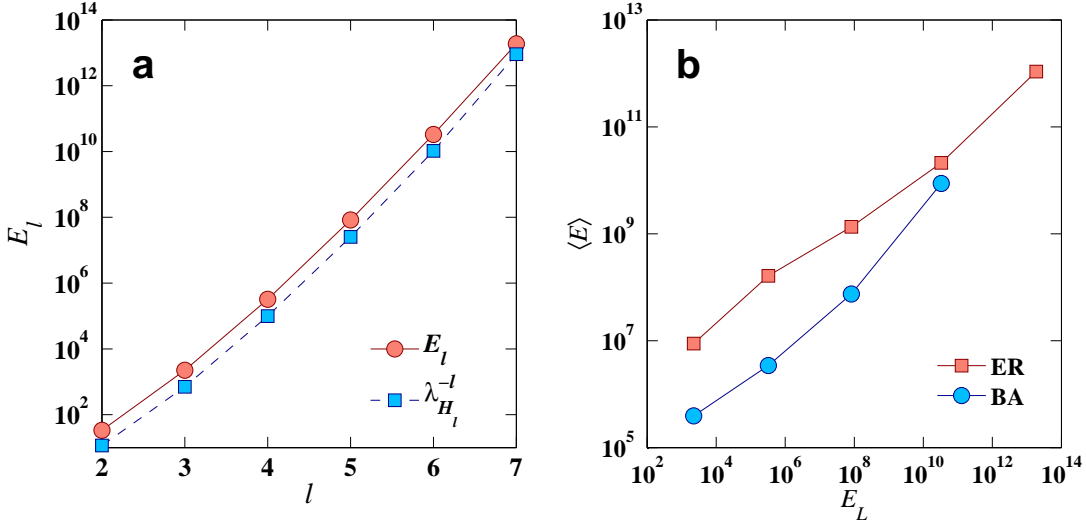}}
\caption{\textbf{Relationship among 1D chain energy, the smallest
eigenvalue of ${H}$-matrix, and network control energy}.
(a) For an 1D chain network of length $l$, energy
$E_l$ and $\lambda_{{H}_l}^{-1}$ versus $l$, providing support
for the analytic result Eq.~(\ref{eq:eigen_H}). (b) Correlation
between $\langle E\rangle$, the average of control energy for
networks with the same LCC length, and $E_L$, the energy of a LCC
with length $D_{\text{C}}=L$ ($L=3$, 4, 5, 6, and 7 for ER and $L=3$, 4, 5,
and 6 for BA networks), calculated from ensembles of $10000$ networks.}
\label{fig:eigen_H}
\end{figure}

We develop a physical understanding of the large control energy required and also
the algebraic scaling behavior in the energy distribution. To gain insights,
we first consider a simple model: an unidirectional, one-dimensional (1D) string
network, for which an analytic estimate of the control energy can be obtained
(see Appendix C) as
\begin{equation} \label{eq:eigen_H}
E_l \approx \lambda_{{H}_l}^{-1},
\end{equation}
where $E_l$ denotes the energy required to control a 1D string of length
$l$ (the number of nodes on the string) and $\lambda_{{H}_l}$ is the smallest
eigenvalue of the underlying ${H}$-matrix, denoted by ${H}_l$, which is
related to the Gramian matrix by $H\equiv e^{-At_\text{f}}W e^{-A^Tt_\text{f}}$.
The condition number of the 1D chain system increases exponentially with
its length. For example, the value of $C_W$ of a chain of length larger
than $7$ has already exceeded $\bar{C}_W = 10^{12}$. This indicates that, even for
a simple 1D chain network, the energy required for control tends to
increase exponentially with the chain length. Numerical verification of
Eq.~(\ref{eq:eigen_H}) is presented in Figs.~\ref{fig:eigen_H}(a).
Although Eq.~(\ref{eq:eigen_H}) is obtained for a simple 1D chain network,
we find numerically that it holds for random and scale-free network topologies
(See Fig.~A5 in Appendix D).

The relation Eq.~(\ref{eq:eigen_H}) and Fig.~\ref{fig:eigen_H}(a) provide an intuitive explanation for our finding that applying control signals to the
minimum set of driver nodes calculated from the structural controllability
theory typically requires enormously large energies. Under the theory of
structural controllability, a network is deemed more structurally controllable
if $N_\text{D}$ is smaller~\cite{LSB:2011}. However, as the number of driver nodes
is reduced, the length of the chain of nodes that each controller drives
on average must increase, leading to an exponential growth in the control
energy. In the ``optimal'' case of structural controllability where
$N_\text{D} = 1$ is achieved, the length of the control chain will be maximized,
leading to unrealistically large control energy that prevents us from
achieving actual control of the system.

\begin{figure*}[ht]
\begin{center}
\centerline{\includegraphics[width=1\textwidth]{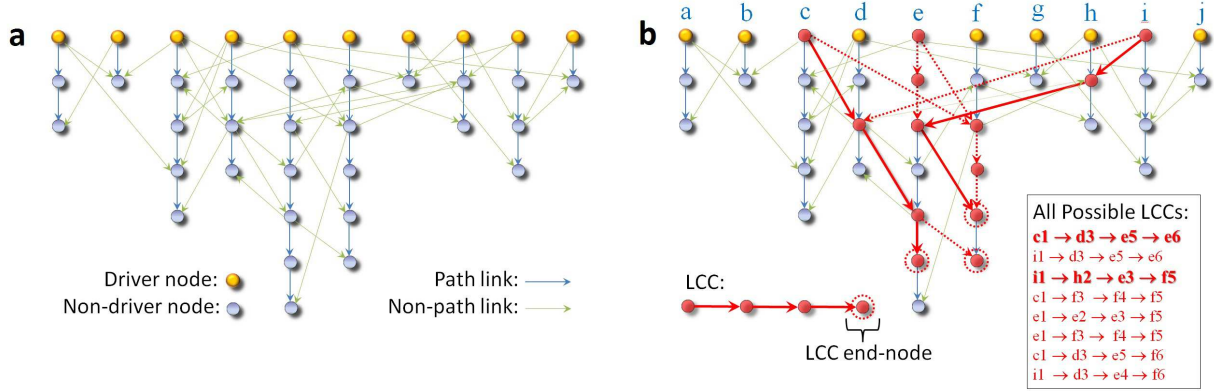}}
\caption{\textbf{Schematic illustration of various concepts to
characterize and understand the practical controllability of a network.}
(a) Control-signal paths (CSPs) of a random network obtained from maximum
matching in structural controllability theory, where a control signal
enters a CSP via the corresponding driver node (yellow), the starting
node of the path, and goes through each matched node (blue) along the
path. In this example, the network has $n_{\text{D}} = 10$ CSPs of
different lengths. Non-path links, links that are irrelevant to
matching, are displayed in green. (d) All possible LCCs in the network.
Typically there are multiple LCCs of the same length. In this example,
the length of the LCCs is 4, which is defined as the control diameter of
the network. Two LCCs sharing no common nodes are marked by red nodes and
solid red arrows. Links belonging to other LCCs are marked by red dashed arrows.
CSPs are denoted using letters $a$ to $j$ from the left to the right. Each
node is specified using its path number and its position along the path
sequentially from top to bottom. For example, node $e1$ is the
driver node of path $e$, and node $h2$ is the node right after the driver
node on path $h$. The two LCCs with solid arrows are listed in bold. Eight LCCs
in the network converge to only three end-nodes, $e6$, $f5$, and $f6$ (marked
by red dashed circles), leading to LCC degeneracy $m=3$. The control energy
is determined by any randomly chosen $m$ LCCs among all existent ones.}
\label{fig:path_chain_LCC}
\end{center}
\end{figure*}

The idea of exploiting the length of control chain can also be used to
explain the algebraic scaling behavior in the energy consumption. As discussed,
identifying maximum matching so that the network is deemed structurally
controllable is independent of the control energy.
However, when maximum matching is found, we
can divide the whole network into $N_\text{D}$ control signal paths (CSPs),
each being a unidirectional 1D string led by a driver node that passes the
control signal onto every node along the path, as illustrated by the vertical
paths in Fig.~\ref{fig:path_chain_LCC}(a). CSPs thus provide a picture
indicating how the signals from the $N_\text{D}$ external control inputs reach
every node in the network to ensure full control (in the sense of structural
controllability).

We can distinguish two types of links: one along and another between the CSPs,
as shown in Fig.~\ref{fig:path_chain_LCC}(a). It may seem that the latter
class are less important as the control signal and energy flow along the
former set of links. However, due to coupling, a node's dynamical state
will affect all its nearest neighbors' states which, in turn, will affect the
states of their neighbors, so on, and vice versa. In principle, any driver
node connects with nodes both along and outside its CSP. Correspondingly,
an arbitrary node in the network is influenced by every driver node, directly
through the CSP to which it belongs, or indirectly through the CSPs that it does
not sit on. Intuitively, the ability of a driver node to influence a node becomes
weaker as the distance between them is increased. In order to control a distant
node, exponentially increased energy from the driver is needed. The chain
starting from a driver node and ending at a non-driver node along their
shortest path is effectively a control chain. We can define the length of the
longest control chain (LCC), $D_{\text{C}}$, as the {\em control diameter} of
the network, as shown in Fig.~\ref{fig:path_chain_LCC}(b). There can be multiple
LCCs. The node at the end of a LCC is most difficult to be controlled in the sense
that the largest amount of control energy is required. The number $m$ of such
end nodes dictates the degeneracy (multiplicity) of LCCs. An
example is shown in Fig.~\ref{fig:path_chain_LCC}(b), where we see that, although
there can be multiple LCCs, the ends of them converge to only three nodes,
leading to $m = 3$. Since the energy required to control a 1D chain grows
exponentially with its length in such a way that even one unit of increase in
the length can amplify the energy by several orders of magnitude
[Fig.~\ref{fig:eigen_H}(a)], the energy associated with any chain shorter than
the LCC can typically be several orders of magnitude smaller than that with the
LCC. Thus, the total energy is dominated by the LCCs. Due to the low value
of typical $m$ (see Fig.~A6 in Appendix E1), a single LCC essentially dictates
the energy magnitude of the whole system. As shown in Fig.~\ref{fig:eigen_H}(b),
actual network control energy shares strong positive correlation and similar magnitude
to the LCC energy $E_L$, defined as the energy of a LCC of the corresponding network,
especially for networks with long LCCs. Intuitively, the probability to form
long LCCs is small. Accordingly, a longer LCC tends to have smaller value of
degeneracy $m$. As a result, the longest LCCs have almost no degeneracy ($m=1$)
so that they effectively rule the control energy of the whole network (see Appendix E1).

The construction in Fig.~\ref{fig:path_chain_LCC} thus provides a structural
profile to estimate the control energy. In particular, a network can be viewed
as consisting of a set of structural elements, the control chains,  interacting with each other via the
links among them, and interactions among these basic structural elements usually play an
important role in determining the properties of a physical system. Hence, the
total energy $E$ required has two components: $E_1$, the sum of energies
associated with all control chains, and $E_2$, the interaction energies among
the chains. Observed from Fig.~\ref{fig:eigen_H}(b), $E_2$ is important
for networks with short LCCs (higher $m$ values). The energy scaling relation
shown in Fig.~\ref{fig:energy_distri} can be derived by devising appropriate
models to analyze the contributions from the two components. We have developed
two such models. The first is the \emph{LCC-skeleton} model, which only takes
$E_1$ into account and provides an analytic estimate of the control energy
distribution function as well as the scaling exponent. The second is the
\emph{double-chain interaction} model, in which a system consisting only
two interacting control chains captures the key features of the entire network by characterizing the essential effect of interaction energy among the structural elements.
These two models combined serve as a framework to determine the energy profile
associated with controlling a complex networked system, providing a deep
understanding of practical controllability (see Appendix E for
details of the two models).

\begin{figure}
\begin{center}
\epsfig{figure=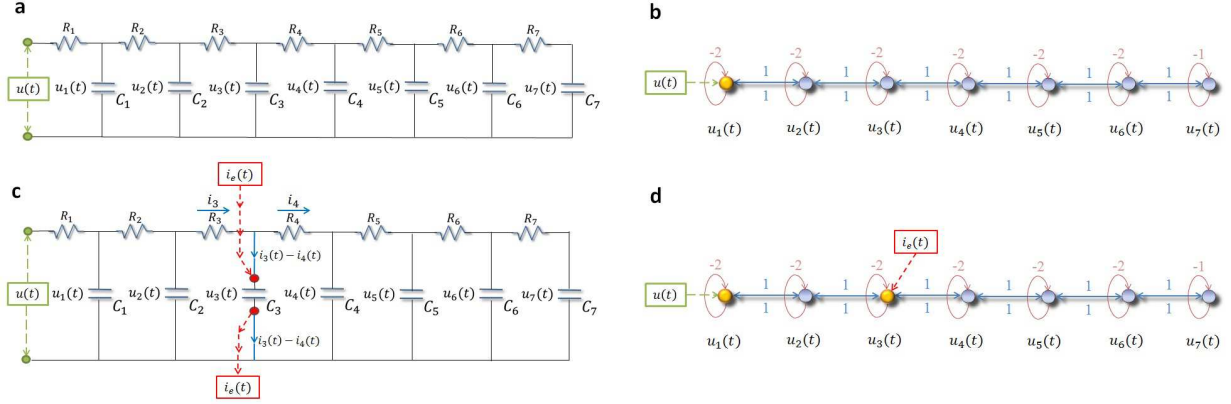,width=\linewidth}
\caption{\textbf{Cascade parallel R-C circuit and its corresponding
network presentation.}
(a) A cascade parallel R-C circuit with $L=7$ resistors ($R_1$, $R_2$,
$\ldots$, and $R_L$, each of resistance $1 \Omega$) and $7$ capacitors
($C_1$, $C_2$, $\ldots$, and $C_L$, each of capacitance $1 \mbox{F}$).
External voltage input $u(t)$ is applied onto the left side of the
circuit, and the voltage of capacitor $C_i$ is $u_i(t) (1\leq i \leq L)$.
(b) Network representation of the circuit in (a) as a bidirectional 1D
chain network of seven nodes, where the external voltage input $u(t)$
is injected into node $1$ (yellow driver node, the controller). The
dynamical state of node $i$ is described by the voltage on its capacitor,
$u_i(t)$. Links (blue) between nodes are bidirectional and have uniform
weight $1$ in either direction. Each node has a self-link (red) of weight
$-2$, except the ending node (node $7$) whose self-link has weight $-1$.
(c) The circuit network in (b) with an extra external current input
$i_{\text{e}}(t)$ into the capacitor $C_3$, where $i_3$ and $i_4$ denote the
currents through resistors $R_3$ and $R_4$, respectively. In the absence
of the extra current input, $i_3(t) - i_4(t)$ is the current through the
branch of $C_3$. (d) The extra external current input $i_{\text{e}}(t)$ serves as
a redundant control input injected into node $3$ of the network in (b).
Now there are two driver nodes (yellow) in the network, nodes $1$ and $3$.}
\label{fig:circuit_A_new}
\end{center}
\end{figure}

\begin{figure}
\begin{center}
\epsfig{figure=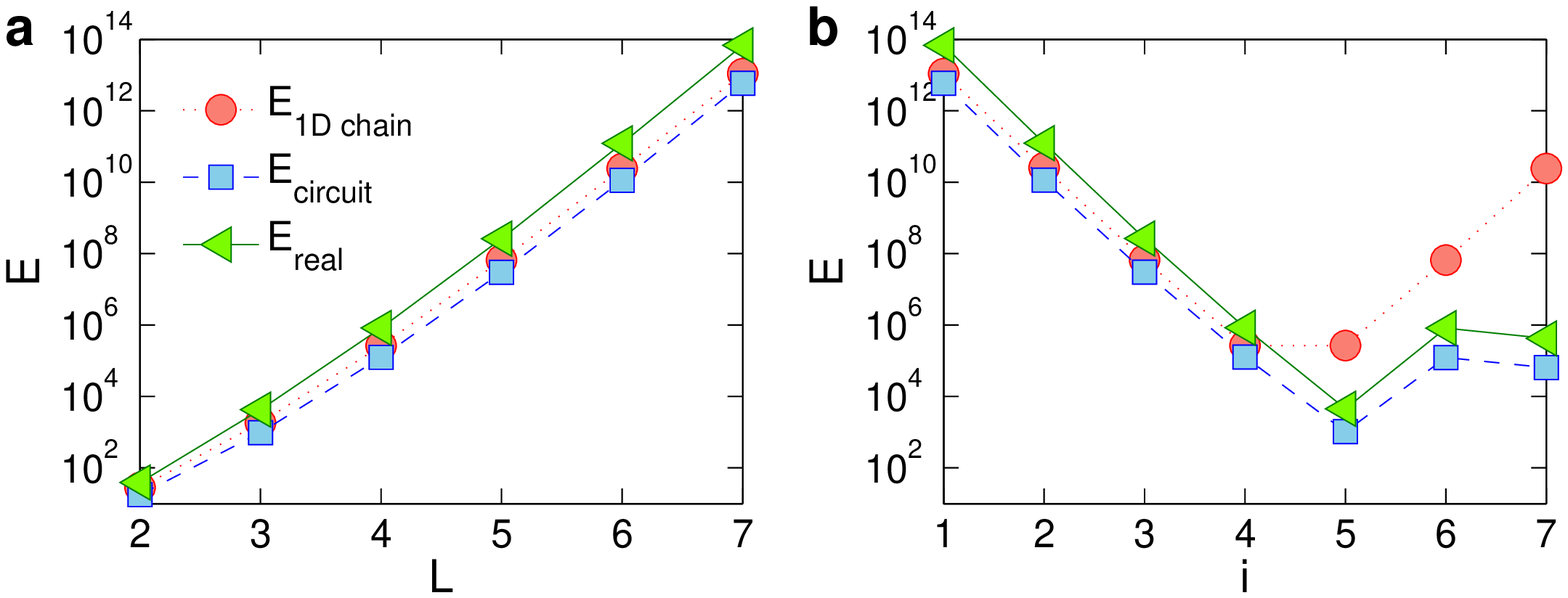,width=\linewidth}
\caption{\textbf{Control energy and optimization for 1D chain and cascade parallel R-C circuit.}
(a) Energy required for controlling a unidirectional
chain (red) and the corresponding circuit (blue) as well as the dissipated
energy of the circuit calculated from Eq.~(\ref{eq:E_real}) versus the
chain length $L$. (b) Control and dissipated energies in the presence
of a redundant control signal to node $i$ ($i > 1$), which breaks the
chain into two subchains of lengths $i$ and $L-i$, respectively.}
\label{fig:circuit_control_perturbation}
\end{center}
\end{figure}

\section{Control of real-world networks}

\subsection{Control of an electrical circuit network}

To further test the concept and framework of practical controllability, we
consider a real one-dimensional cascade parallel R-C circuit network, as 
schematically illustrated in Fig.~\ref{fig:circuit_A_new}(a). The network 
can be represented by a bidirectional 1D chain with self-loops for all the 
nodes, as shown in Fig.~\ref{fig:circuit_A_new}(b). The network size can be 
enlarged, say by one unit, by attaching an additional branch of resistor 
and capacitor at the right end of the circuit. The state $u_i(t)$ of node 
$i$ at time $t$ is the voltage of capacitor $i$, and the input voltage $u(t)$ 
represents the control signal. The purpose of control is to drive the voltages 
of the capacitors from a set of values to another within time $t_{\text{f}}$ 
through the input voltage $u(t)$. The control energy can then be calculated
by Eq.~(\ref{eq:energy}). The actual energy dissipated in the circuit during 
the control process is given by
\begin{equation} \label{eq:E_real}
E_{\text{real}} = \int_{0}^{t_{\text{f}}} U(t) \cdot I(t) dt,
\end{equation}
where $U(t) \equiv u(t)$ and $I(t)$ are the input voltage and current at 
time $t$, and $E_{\text{real}}$ is in units of Joule. By making the circuit 
equivalent to a 1D chain network, we have three types of energy: the control 
energy of the actual circuit calculated from Eq.~(\ref{eq:energy}), 
the dissipated energy of the circuit from Eq.~(\ref{eq:E_real}), and the 
control energy of the 1D equivalent network. 
Figure~\ref{fig:circuit_control_perturbation}(a) shows 
that the control energy and the dissipated energy of the circuit do not 
differ substantially from the energy calculated from unidirectional 1D chain. 
Among the three types of energy, the energy cost associated with the control 
process, as calculated from Eq.~(\ref{eq:E_real}), is maximal. 

\begin{figure}
\begin{center}
\epsfig{figure=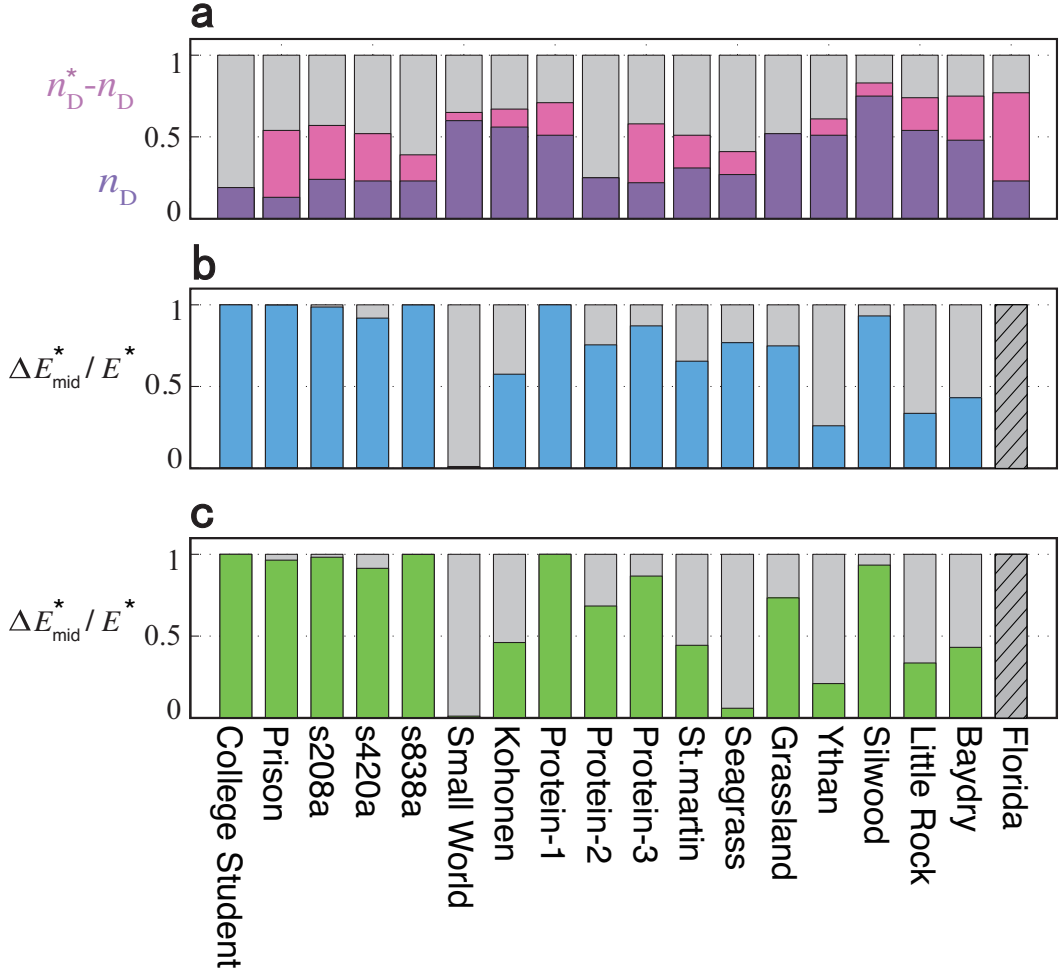,width=\textwidth}
\caption{\textbf{Augmented control inputs and energy optimization.}
(a) Densities of the original driver nodes $n_{\text{D}}$ (purple) and of the 
augmented controls  $n^\star_{\text{D}}-n_{\text{D}}$ (pink). (b) Normalized 
energy reduction 
$\Delta E^\star_{\text{mid}}/E^\star = (E^\star-E^\star_{\text{mid}})/E^\star$ 
(blue) when an additional control signal is added to the middle of each LCC 
[strategy (I)]. (c) Normalized energy reduction 
$\Delta E^\star_{\text{end}}/E^\star = (E^\star-E^\star_{\text{end}})/E^\star$ 
(green) when an additional control signal is added to the end node of each LCC 
[strategy (II)]. For the blue bars (or green bars), the optimized control 
energy $E\star_{\text{mid}}$ (or $E\star_{\text{end}}$) is several orders of 
magnitude smaller than $E\star$. The bars with more gray portion the blue (or 
green) potions are for the networks with relatively low values of $E\star$ 
and $D_{\text{C}}$ (see Table A2 in Appendix F), for which energy optimization
is not necessary. The bar corresponding to ``Florida'' (the most right) is 
striped due to the fact that this network is not controllable (i.e., with
divergent energy) even if $M\star$ augmented control inputs are added. }
\label{fig:optimization_bar}
\end{center}
\end{figure}

\subsection{Strategies to balance control energy and extra inputs}


Our finding of the LCC structure associated with the control and the
exponential growth of energy with the length of LCCs suggest a method to
reduce the energy significantly.
Since the key topological structure that determines the control
energy is LCCs, one possible approach is to reduce the length of all the LCCs
embedded in a network by making structural perturbations to the network. This,
however, will inevitably modify the network structure, which may not always be
practically viable. Is it possible to reduce the control energy without having
to change the network structure? One intuitive method is to apply additional
controllers beyond those calculated from the structural-controllability theory,
which we name as {\em redundant} controllers. A straightforward solution is 
to add some redundant control signals along the LCCs. To gain insights, 
we consider a unidirectional 1D chain and add a
redundant control input at the $i$th node. As shown in
Fig.~\ref{fig:circuit_control_perturbation}(b), the magnitude of control
energy is reduced dramatically. The optimal location to place the extra
control should be near the middle of the chain so as to minimize the length of
LCCs using a minimal number of redundant control signals. As can be
seen from Fig.~\ref{fig:circuit_control_perturbation}(b), this simple strategy
of adding one redundant control signal can reduce the required energy by nearly
seven orders of magnitude! More specifically, the redundant control signal to
node $i$ breaks a chain of length $L$ into two shorter subchains: one of
length $i-1$ and another of length $L-i+1$. Roughly, the control energy
is the sum of energies required to control the two shorter components,
which is dominated by energy associated with the longer component owing
to the exponential dependence of the energy on the chain length. By
choosing $i$ around $L/2$, the length of the longer part is minimized.
For the circuit network in Fig.~\ref{fig:circuit_A_new}, the redundant
control input can be realized by inducing external current input into
a capacitor. As shown in Fig.~\ref{fig:circuit_A_new}(b), a reduction
in energy of nearly $10$ orders of magnitude is achieved. Applying a
single redundant control input can thus be an extremely efficient strategy
to reduce the required control energy for the one-dimensional chain network.

Due to the fact that there can be multiple LCCs converge at the same end 
node, applying a control signal to each of the $m$ nodes that all LCCs
converge into is another strategy that reduces the total number of redundant 
controls, while also significantly shrinks the control energy. (Detailed 
demonstrations of the enhancement strategies for physical or modeled
networks and an implementation example on a circuit system are presented 
in Appendix G.)

\subsection{Control of real-world networks}

Can real-world complex networks be actually controlled? In Ref.~\cite{LSB:2011},
the structural controllability of a large number of real-world networks
were investigated, with the conclusion that optimal control of most of the
networks can be achieved with only a few control signals. We investigate
control energies of the same set of real-world networks (see Table A1 in
Appendix F for network details) and find that, when optimal
control is applied according to maximum matching, most of the networks
require realistically high energies. In fact, $15$ out of the $18$ networks are
practically uncontrollable. The main reason lies in the large LCCs of most
of these networks. Another factor is that there are subgraphs that are not
connected with each other and/or a large number of topological motifs such
as loops, self-loops, or bidirectional edges. More strikingly, even with
unlimited energy supply, the number of driver nodes as determined by the
maximum matching algorithm from the structural controllability theory is
generally insufficient to fully control the whole system, where there exists
a number $M^\star$ of nodes that never converge to their target states. These
observations lead to the speculation that, in order to fully control a realistic
network, more driver nodes are needed than those identified by the structural
controllability theory. That is, more independent control signals are needed
than those determined by maximum matching to drive all nodes in the network to
their target states. The $M^\star$ uncontrollable nodes are thus the required
augmented set of driver nodes, each with an external control input. In total, $N_D^\star = N_D + M^\star$ driver nodes need to be deployed to gain full control of the system [see Fig.~\ref{fig:optimization_bar}(a) for $n_D$ and $n_D^\star = N_D^\star/N$ for the $18$ real-world networks]. Applying control
signals to the nodes as determined by maximum matching and to the augmented
driver nodes, we find that $17$ out $18$ real-world networks become practically
controllable (see Table A2 in Appendix F).

We also test the enhancement strategies using the $18$ real-world networks, 
with the result that their practical controllability can be markedly enhanced 
(especially for those with large control diameters), as shown 
Figs.~\ref{fig:optimization_bar}(b) and (c). We see that, for each of the 
real-world networks with unrealistically large energy requirement (see 
Table A2 in Appendix F), the optimized control energy $E^\star_{\text{mid}}$ 
(or $E^\star_{\text{end}}$) is several orders of magnitude smaller than the 
value of the original energy $E^\star$ (the control energy with $M^\star$ 
augmented driver nodes but without any redundant control input). This 
indicates the effectiveness of our optimization strategies. (In fact, 
strategy (I) works better than (II) in most cases.) For the networks with
small control diameters, even without applying any enhancement strategy 
the control energies required are already much smaller than those 
for the other networks. For these networks energy optimization is 
practically unnecessary (see also Table A2 in Appendix F).

We also find that increasing the control time $t_{\text{f}}$ can reduce 
the control energy so as to enhance the network's practical controllability.

\section{Conclusions and Discussions}

As stated in Ref.~\cite{LSB:2011}, the ultimate proof that one understands a
complex network completely lies in one's ability to control it. We discover
a paradox arising from controlling complex networks with respect to control
energy and the number of external input signals. To resolve the paradox,
we focus on the situation where the structural-controllability theory yields
a minimum number of external input signals required for full control of the
network, and determine whether in these situations the control energy is
affordable so as to realize actual control. Our systematic computations and
analysis reveal a rather unexpected phenomenon: due to the singular nature of
the control Gramian matrix, in the parameter regimes where optimal structural
controllability is achieved in the sense that the number of driver nodes is
minimized, energy consumption can be unbearably large. To obtain a more systematic
understanding, we identify the fundamental structures in a network under the
action of control signals, the longest control chains (LCCs), and argue that
they essentially determine the control energy. We articulate and validate that
the required energy increases exponentially with the length of the LCCs. In
situations where the required number of controllers is few as determined by the
structural controllability theory, the length of LCCs tends to be long,
leading to practically divergent control energy. Another finding is that,
for minimum input signals, the required energy exhibits a robust algebraic
scaling behavior, which can be explained by analyzable models constructed
based on interacting LCCs. The discovery of the LCCs associated with controlling
complex networks leads naturally to a simple method to resolve the paradox:
increasing the number of controllers by
placing extra control signals (beyond the number determined by the
structural-controllability theory) along the LCCs. Indeed, test of a large
number of real-world networks shows that, while they are structurally
controllable~\cite{LSB:2011}, most of them exhibit enormous energy consumption.
They can actually be controlled by placing more drivers than determined by
the structural-controllability theory at proper locations along the LCCs.

Our work indicates that the difficulty of achieving actual control of
complex networks associated with even linear dynamics is beyond the current
knowledge in the field of network control. Although the controllability
theory offers a theoretically justified framework to guide us to apply
external inputs on a minimum set of driver nodes, when we implement
control to steer a system to a desired state, the energy consumption is likely
to be too large to be affordable. This finding suggests that, to achieve
control of a complex networked system, the existing controllability framework
merely offers a necessary rather than a practically feasible condition to
assure actual control. We thus demand a more comprehensive and practically
useful theoretical framework for addressing the extremely important issue
of controlling complex networks. However, it is difficult to develop such
a framework at the present and we do not even know if a mathematically
justified theory is available based on the current knowledge. Another
issue is that for general networked nonlinear systems, we continue to
lack the necessary condition based on the present controllability framework, as
well as an understanding of required control energy. So far, we still know too
little about controlling complex networked systems, and further effort is needed
to address this challenging but greatly important problem shared by a
wide range of fields.

\section*{acknowledgments}

The first two authors contributed equally. We thank Dr. H. Liu for tremendous
help with Appendix C. This work was supported by ARO under Grant 
No.~W911NF-14-1-0504. W.-X.W. was supported by NSFC under Grant No.~11105011.

\newpage
\small
\renewcommand\theequation{A\arabic{equation}}
\renewcommand\thefigure{A\arabic{figure}}
\renewcommand\thetable{A\arabic{table}}
\setcounter{equation}{0}
\setcounter{figure}{0}
\setcounter{table}{0}
\newcommand{\dif}{\,\mathrm{d}}
\newcommand{\pd}[2]{\displaystyle\frac{\partial#1}{\partial#2}}
\newcommand{\ppd}[2]{\displaystyle\frac{\partial^2#1}{\partial#2^2}}
\newcommand{\fd}[2]{\displaystyle\frac{\dif#1}{\dif#2}}
\newcommand{\ffd}[2]{\displaystyle\frac{\dif^2#1}{\dif#2^2}}
\captionsetup[figure]{labelfont=bf, name=Figure, labelsep=vline}

\section*{Appendix A: Analytical calculation of the in- and out-degree
distributions}

The in- and out-degree distributions of a directed complex network
under connection bias probability $\lambda \equiv P_{\text{b}}$
can be obtained analytically.

Defining $k_{\text{S}}$ and $k_{\text{L}}$ to be the numbers of nodes in the
neighborhood of a node with degree $k$, whose degrees are smaller
or larger than than $k$, respectively, we have
\begin{equation}
k_{\text{S}} = k\sum_{k'=k_{\text{min}}}^{k}P(k'|k)
= \frac{k}{\langle k \rangle} \sum_{k_{\text{min}}}^{k} k' P(k)
= \frac{k}{\langle k \rangle} \int_{k_{\text{min}}}^{k}
k'\cdot Ck'^{-\gamma}\dif k' = \frac{Ck}{\langle k
\rangle}\int_{k_{\text{min}}}^{k} k'^{1-\gamma}\dif k',
\end{equation}
and
\begin{equation}
k_{\text{L}} = k\sum_{k'=k}^{k_{\text{max}}}P(k'|k)
=\frac{Ck}{\langle k \rangle}\int_{k}^{k_{\text{max}}} k'^{1-\gamma}\dif k'
\stackrel{\gamma >2}{=} \frac{Ck^{3-\gamma}}{\langle
k\rangle (\gamma - 2)} = Ak^{3-\gamma},
\end{equation}
where
\begin{equation}
A=\frac{C}{\langle k\rangle (\gamma - 2)}.
\end{equation}
Therefore,
\begin{equation}
k_{\text{S}} = k - k_{\text{L}}.
\end{equation}
We then have
\begin{equation}
k_{\text{out}} = \lambda k_{\text{S}} + (1-\lambda)k_{\text{L}} =
\lambda(k-k_{\text{L}}) + (1-\lambda)k_{\text{L}} = (1-2\lambda)k_{\text{L}}+
\lambda k = (1-2\lambda)Ak^{3-\gamma} + \lambda k,
\end{equation}
and
\begin{equation}
\nonumber k_{\text{in}} = (1-\lambda)k_{\text{S}} + \lambda k_{\text{L}} =
(1-\lambda)k + (2\lambda -1)Ak^{3-\gamma}.
\end{equation}
The quantities $P(k_{\text{out}})$ and $P(k_{\text{in}})$ can be derived from
\begin{equation}
P(k_{\text{out}})\dif k_{\text{out}} = P(k)\dif k \ \ \mbox{and} \ \
P(k_{\text{in}})\dif k_{\text{in}} = P(k)\dif k,
\end{equation}
which yield
\begin{equation}
P(k_{\text{out}}) = P(k)\frac{\dif k}{\dif k_{\text{out}}} \ \ \mbox{and} \ \
P(k_{\text{in}}) = P(k)\frac{\dif k}{\dif k_{\text{in}}}.
\end{equation}
Setting $k=f_1^{-1}(k_{\text{out}})$ and
$k=f_2^{-1}(k_{\text{in}})$, we can obtain the distributions.

Using
\begin{equation}
\nonumber k_{\text{out}} = f_1(k) \ \ \mbox{and} \ \
k_{\text{in}} = f_2(k),
\end{equation}
we obtain
\begin{equation}
P(k_{\text{out}}) = P(k) \frac{1}{\frac{\dif k_{\text{out}}}{\dif k}} =
P(f_1^{-1}(k_{\text{out}}))\frac{1}{f_1'(f_1^{-1}(k_{\text{out}}))},
\end{equation}
and
\begin{equation}
P(k_{\text{in}}) = P(f_2^{-1}(k_{\text{in}}))\frac{1}{f_2'(f_2^{-1}(k_{\text{in}}))}.
\end{equation}
In general, it is difficult to obtain an explicit expression. However, for
some specific values of $\lambda$ or $\gamma$, analytical results
are available.\\

\textbf{Case I}: $\lambda = 0.5$. \\

In this case, we have
\begin{equation}
k_{\text{out}} = k_{\text{in}} = \frac{k}{2},
\end{equation}
and
\begin{equation}
k =2k_{\text{out}} = 2k_{\text{in}}.
\end{equation}
Thus
\begin{equation}
\nonumber P(k_{\text{out}}) = 2P(2k_{\text{out}}) =
2C(2k_{\text{out}})^{-\gamma} = 2^{1-\gamma}C k_{\text{out}}^{-\gamma}.
\end{equation}
Akin to $P(k_{\text{out}})$, we have
\begin{eqnarray}
\nonumber P(k_{\text{in}}) = P(k_{\text{out}}) = 2^{1-\gamma}C k_{\text{in}}^{-\gamma}.
\end{eqnarray}
\\ \indent
\textbf{Case II}: $\lambda = 0$. \\
\begin{eqnarray}
k_{\text{out}} = Ak^{3-\gamma} \Rightarrow
k=\frac{k_{\text{out}}}{A}^{\frac{1}{3-\gamma}}.
\end{eqnarray}
We then have
\begin{equation}
\nonumber P(k_{\text{out}}) =
C(\frac{k_{\text{out}}}{A})^{\frac{1}{3-\gamma}(-\gamma)}\frac{\dif k}{\dif k_{\text{out}}}
=
C(\frac{k_{\text{out}}}{A})^{\frac{\gamma}{\gamma-3}}\frac{1}{3-\gamma}\bigg(
\frac{k_{\text{out}}}{A}\bigg)^{\frac{1}{3-\gamma}-1}\frac{1}{A} =
\frac{C}{3-\gamma}A^{\frac{1}{3-\gamma}}k_{\text{out}}^{\frac{2}{\gamma -
3}}.
\end{equation}

\textbf{Case III}: $\lambda = 1$. \\

\begin{eqnarray}
k_{\text{in}} = Ak^{3-\gamma} \Rightarrow
k=\frac{k_{\text{in}}}{A}^{\frac{1}{3-\gamma}},
\end{eqnarray}
which yields
\begin{equation}
P(k_{\text{in}}) = P(k_{\text{out}}) =
\frac{C}{3-\gamma}A^{\frac{1}{3-\gamma}}k_{\text{in}}^{\frac{2}{\gamma - 3}}.
\end{equation}

\textbf{Case IV}: $\gamma = 3$. For example, for BA model, we have
\begin{equation}
k_{\text{out}} = (1-2\lambda)A + \lambda k,
\end{equation}
and
\begin{equation}
k=\frac{k_{\text{out}} + (2\lambda -1)A}{\lambda}.
\end{equation}
\begin{equation}
\nonumber P(k_{\text{out}}) = C\bigg( \frac{k_{\text{out}} + (2\lambda
-1)A}{\lambda} \bigg)^{-\gamma}\frac{1}{\lambda} =
C\lambda^{\gamma -1}\bigg[ k_{\text{out}} + (2\lambda -1)\frac{C}{\langle
k \rangle} \bigg]^{-\gamma} = C\lambda^{\gamma -1}\bigg[
k_{\text{out}} + (2\lambda -1)k_{\text{min}} \bigg]^{-\gamma}.
\end{equation}
Similarly, we have
\begin{equation}
k_{\text{in}} = (2\lambda-1)A + (1-\lambda) k,
\end{equation}
and
\begin{equation}
k=\frac{k_{\text{in}} + (1-2\lambda )A}{1-\lambda},
\end{equation}
so
\begin{equation}
P(k_{\text{in}}) = C\bigg[ \frac{k_{\text{in}} + (1-2\lambda
)A}{1-\lambda} \bigg]^{-\gamma}\frac{1}{1-\lambda}
= C(1-\lambda)^{\gamma-1}[ k_{\text{in}} + (1-2\lambda)k_{\text{min}}]^{-\gamma}.
\end{equation}

\newpage

\section*{Appendix B: Additional numerical results}

\begin{figure}
\begin{center}
\epsfig{figure=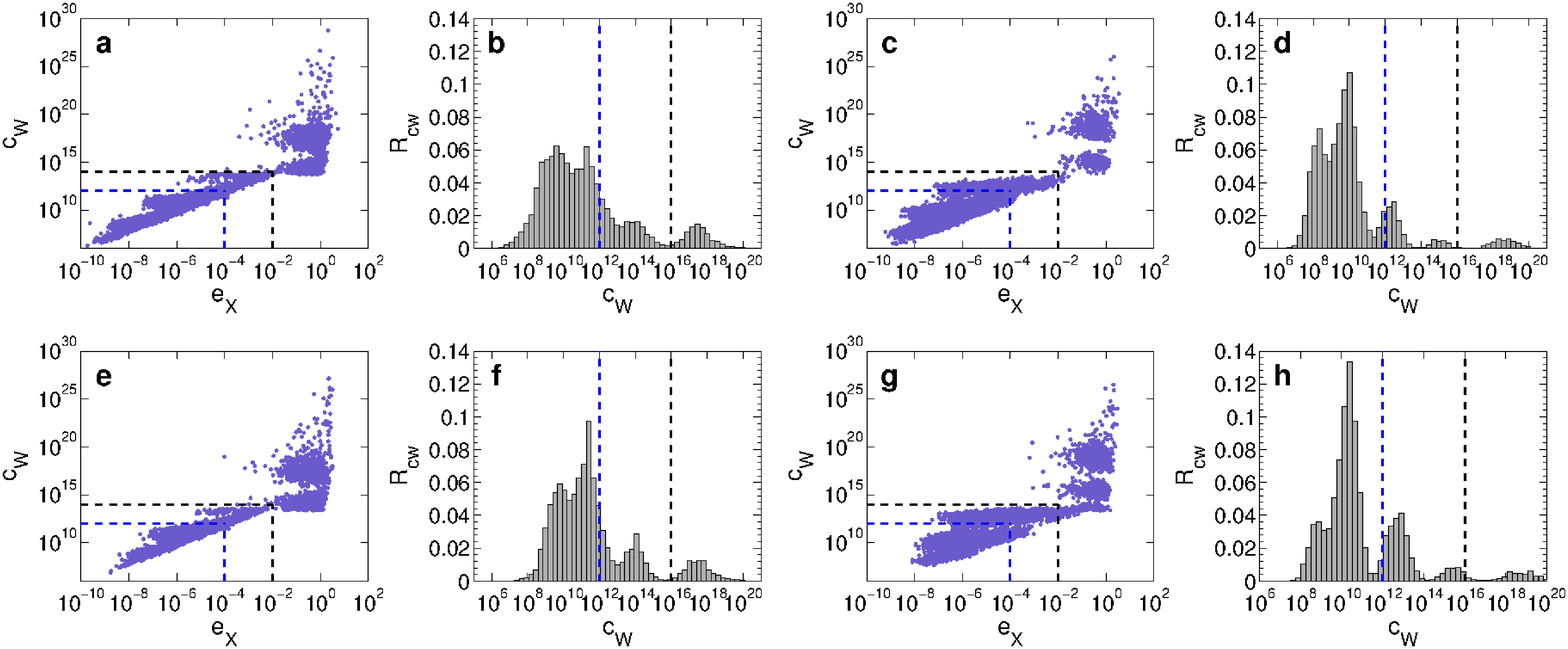,width=\linewidth}
\caption{\textbf{Condition number $C_W$ versus control error $e_X$ for
random and scale-free networks}. Network size is $N=100$ for (a-d) and
$200$ for (e-h), average degree is $\langle {k}\rangle=6$ for ER random
networks [(a),(b),(e), and (f)] and $8$ for BA scale-free networks
[(c), (d), (g), and (h)]. Directional link probability between any
pair of nodes is $P_{\text{b}}=0.1$. Panels (a),(c),(e), and (g) show the scaling
relation between the condition number $C_W$ and control error $e_X$.
Panels (b), (d), (f), and (h) show the fraction $R_{CW}$ of the networks
with a certain $C_W$ number. The scaling relation holds within some
$C_W$-$e_X$ region with boundaries specified as the black dashed lines.
The $e_X$ values are not physically meaningful outside the boundaries
that are defined according to the precision limit of computation. The
thresholds of $C_W$ and $e_X$ used in the computations are $10^{12}$
and $10^{-4}$, respectively, which are indicated as the blue dashed
lines. The threshold values are chosen to lie within the physical
boundaries so that the calculations for all $C_W$ values are meaningful.}
\label{figS:cond_e}
\end{center}
\end{figure}

\paragraph*{B1: Condition number and control error.}
The correlation between the condition number $C_W$ and the control error
$e_X$ is shown in Fig.~\ref{figS:cond_e}. We observe that, within a
certain range of $C_W$, an approximate scaling relation exists between
$C_W$ and $e_X$, shown in panels (a), (c), (e), and (g). However, the
scaling disappears outside the shown $C_W$ range. The reason is that,
outside the $C_W$ range, the Gramian matrix $W$ is ill conditioned,
leading to considerable errors when computing the matrix inverse.
In principle, the scaling regime can be extended with improved
computational precision, but not indefinitely.

\begin{figure}
\begin{center}
\epsfig{figure=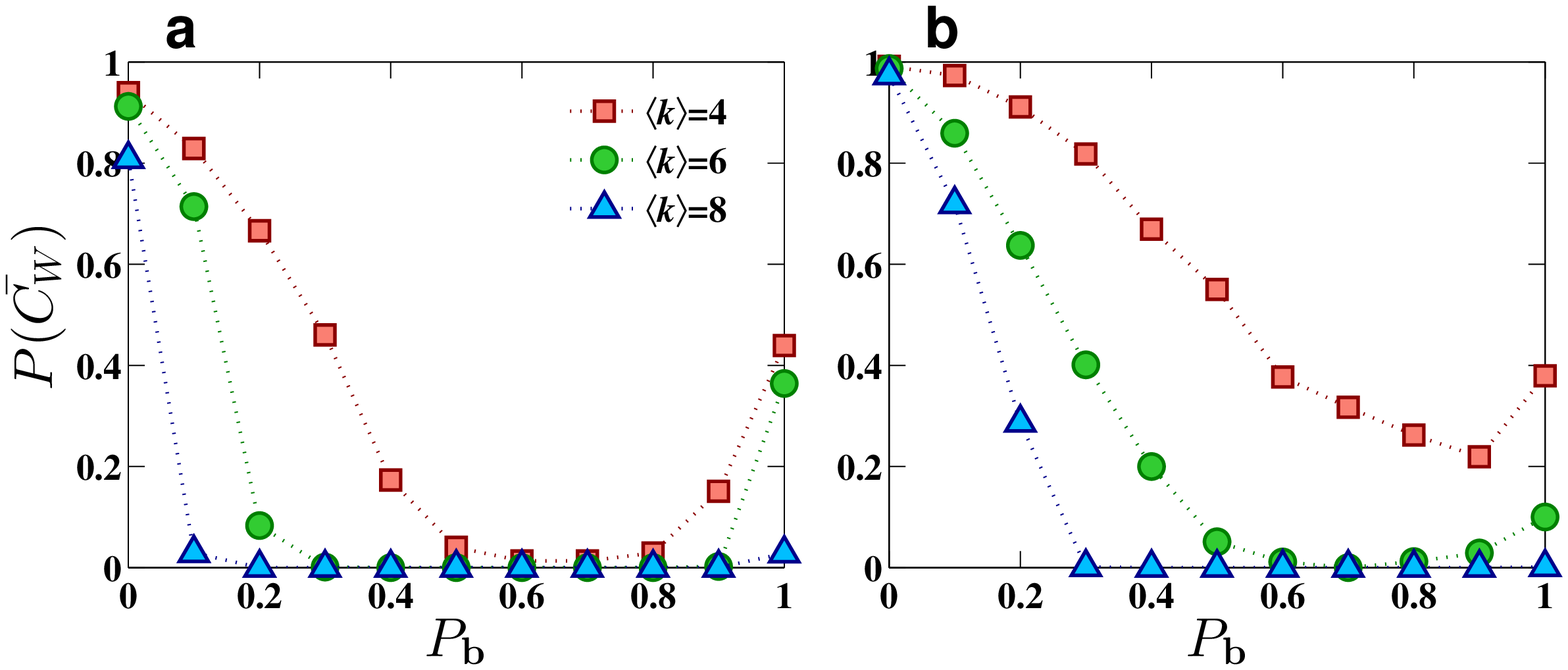,width=\linewidth}
\caption{\textbf{Practical controllability measures for directed networks}.
Measure of the practical controllability, $P(\bar{C}_W)$, for (a) ER random
networks and (b) BA scale-free networks of size $N=200$.}
\label{figS:main}
\end{center}
\end{figure}

\paragraph*{B2: More on structural and practical controllability measures.}
Figure~\ref{figS:main} shows the measure of the structural controllability,
$n_{\text{D}}$, and the measure of the practical controllability, $P(\bar{C_W})$,
versus $P_{\text{b}}$ for ER random and BA scale-free networks of size $N=200$.
We see that the structural and practical controllability cannot be
simultaneously optimized irrespective of the network size.

\begin{figure}
\begin{center}
\epsfig{figure=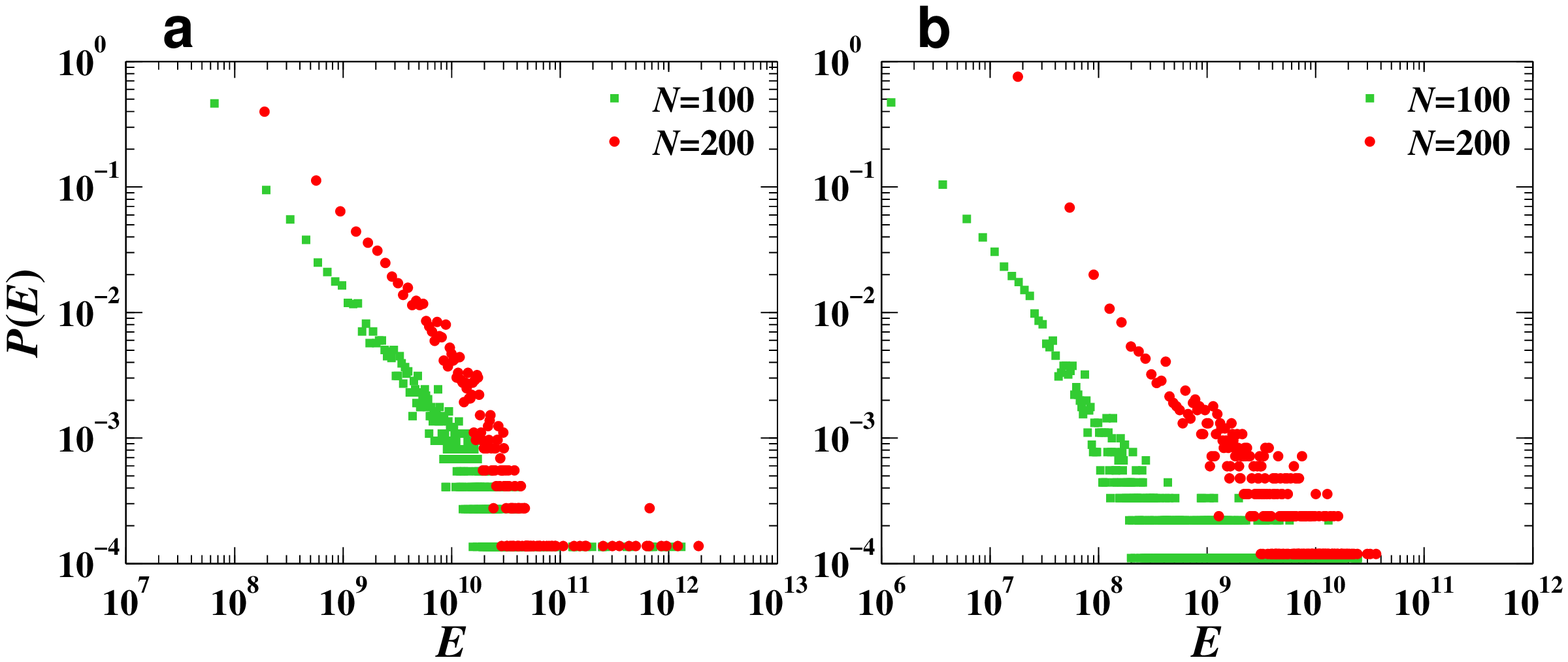,width=\linewidth}
\caption{\textbf{Distributions of control energy for practically controllable
networks under different network sizes}.
Panels (a) and (b) are for random networks with $\langle k\rangle=6$ and
scale-free networks with $\langle k\rangle=8$, respectively, for two values
of the network size ($N = 100$ and $N = 200$), where we set $\bar{C}_W = 10^{12}$
and $t_{\text{f}}=1$. In all cases, we observe an algebraic (power-law) scaling behavior.}
\label{figS:energy_distri}
\end{center}
\end{figure}

\paragraph*{B3: More on control energy power law distribution.}
Figure~\ref{figS:energy_distri} shows the robustness of the control
energy power-law distribution against varying network size for both
random [(a)] and scale-free [(b)] topology.

\clearpage

\newpage
\section*{Appendix C: Control Energy of One-Dimensional String}

\begin{figure}
\begin{center}
\epsfig{figure=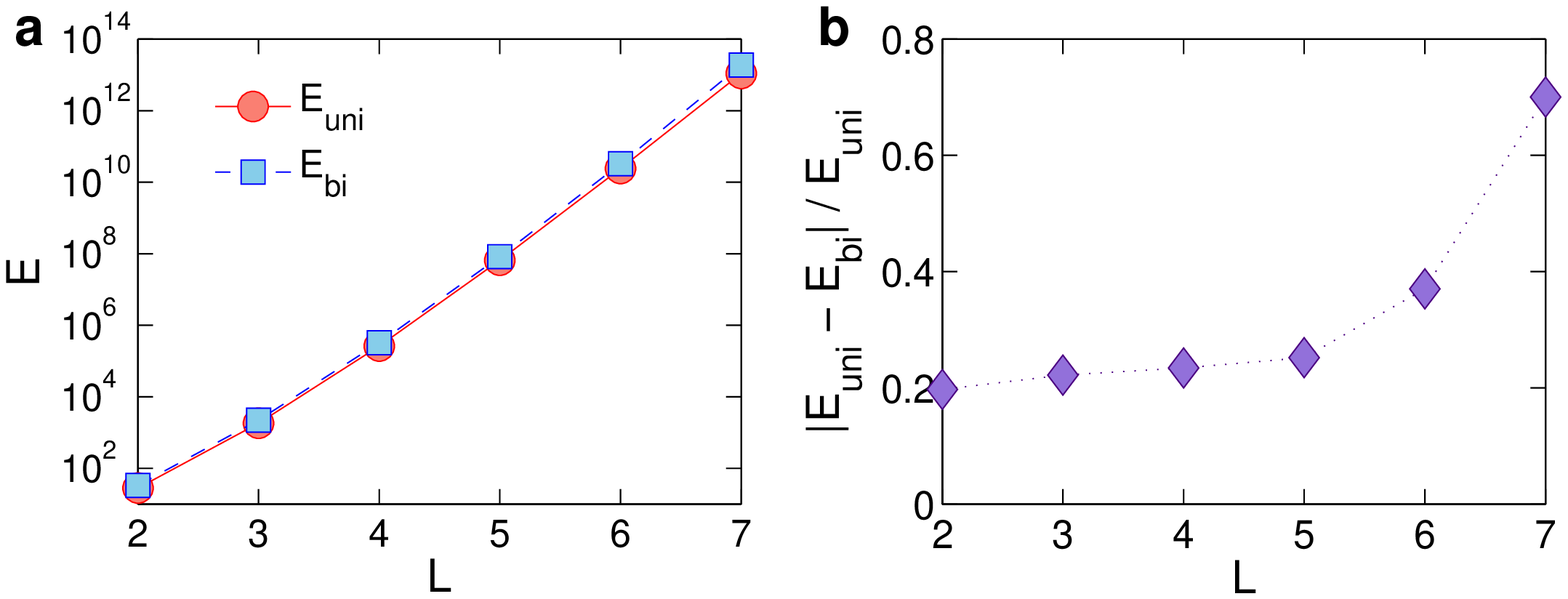,width=\linewidth}
\caption{\textbf{Comparison between energies required to control a
unidirectional and a bidirectional 1D chains:} (a) energies required
to control a unidirectional chain $E_{\text{uni}}$ (red) and a
bidirectional one $E_\text{{bi}}$ (blue) versus chain length $L$,
and (b) the relative energy difference
$|E_\text{{uni}} - E_{\text{bi}}| / E_\text{{uni}}$ versus chain
length $L$.}
\label{fig:bi_uni_directional_chain}
\end{center}
\end{figure}

As shown in Fig.~\ref{fig:bi_uni_directional_chain}, the energy required to
control a unidirectional 1D string nearly overlaps with that of a bidirectional
one with identical weights. In fact, if the chains are not too long, the
relative difference in the energy between the two case are within the
same order of magnitude. Here we provide an analytical calculation of
the control energy for a bidirectional 1D chain network.

The energy $E$ is given by
\begin{equation} \label{eq:E_H}
E(t_{\text{f}})=\mathbf{x}_0^{T}\cdot H^{-1} \cdot \mathbf{x}_0,
\end{equation}
where $H\equiv e^{-At_{\text{f}}}\cdot W \cdot e^{-A^Tt_{\text{f}}}$, $\mathbf{x}_0$ is
the initial state of the network, and $W$ is the Gramian matrix.
Since $H$ is positive definite and symmetric, its inverse $H^{-1}$ can
be decomposed in terms of its eigenvectors as $H^{-1}=Q\Lambda Q^T$,
where $Q=[q_1,q_2,\dots, q_N]$ is composed of the orthonormal eigenvectors
that satisfy $QQ^T=Q^TQ=I$, and
$\Lambda=\mbox{diag}\{\lambda_1,\lambda_2,\dots,\lambda_N\}$
is the diagonal eigenvalue matrix of $H^{-1}$ in descending order.
Numerically, we find that $\lambda_1$ is typically much larger than other
eigenvalues. We thus have
\begin{equation} \label{eq:approx_E1}
E(t_{\text{f}}) = \mathbf{x}_0^TQ\Lambda Q^T\mathbf{x}_0 = \sum_{i=1}^{n}
\lambda_i(q_{i}^{T}\mathbf{x}_0)^2\approx \lambda_1(q_{1}^{T}\mathbf{x}_0)^2.
\end{equation}

Since $\mathbf{x}_0$ can be chosen arbitrarily, we set $\mathbf{x}_0=[1,0,\dots,0]^T$, so
Eq.~(\ref{eq:approx_E1}) becomes
\begin{equation} \label{eq:approx_E2}
E(t_{\text{f}}) \approx \lambda_1(q_{1}^{T}q_{1})=\lambda_1.
\end{equation}

For an undirected network, the adjacency matrix $A$ is positive definite
and symmetric. We can decompose $A$ into the form $A=VSV^{T}$, where the
columns of $V$ constitute the orthonormal eigenvectors of $A$ and
$S=\mbox{diag}\{s_1,s_2,\dots,s_N\}$ is the diagonal eigenvalue matrix of $A$
in descending order. We thus have
$H=e^{-At_{\text{f}}}W e^{-A^{T}t_{\text{f}}}=Ve^{-St_{\text{f}}}V^TWVe^{-St_{\text{f}}}V^T$.
Let
\begin{displaymath}
\Lambda_H=\mbox{diag}\{\lambda_{H_1},\lambda_{H_2},\dots, \lambda_{H_N}\}
=\mbox{diag}\{1/ \lambda_{N},1/ \lambda_{N-1},\dots,1/ \lambda_{1}\}
\end{displaymath}
be the eigenvalue matrix of $H$ in descending order. The energy can thus
be expressed as
\begin{displaymath}
E(t_{\text{f}}) \approx \lambda_1(q_{1}^{T}\mathbf{x}_0)^2=\lambda_{H_N}^{-1}(q_{1}^{T}\mathbf{x}_0)^2.
\end{displaymath}

Letting $\Lambda_W=\mbox{diag}\{\lambda_{W_1},\lambda_{W_2},\dots,\lambda_{W_N}\}$
be the eigenvalue matrix of $W$ in descending order. We can approximate the
eigenvalue of $H$ by $W$, which has been numerically validated:
$\Lambda_H\approx\Lambda_W$. We thus have
\begin{equation} \label{eq:E_eig_W}
\varepsilon\approx \lambda_1(q_{1}^{T}\mathbf{x}_0)^2 =
\lambda_{H_N}^{-1}(q_{1}^{T}\mathbf{x}_0)^2 \approx \lambda_{W_N}^{-1}(q_{1}^{T}\mathbf{x}_0)^2.
\end{equation}

Since orthonormal transform does not alter the eigenvalues of a given matrix,
we have $\Lambda_H=e^{-St_{\text{f}}}\Lambda_We^{-St_{\text{f}}}$.

For an undirected chain, the adjacency matrix is
\begin{equation*}
A = \begin{bmatrix}
 0 & 1 & & & \\
 1 & 0 & \ddots & &\\
 & 1 & \ddots & 1 & \\
 & & \ddots & 0 & 1 \\
 & & & 1 & 0
\end{bmatrix}_{N \times N},
\end{equation*}
control matrix is $B=[1,0,\dots,0]^T$, and eigenvalues and eigenvectors of $A$ are
\begin{align}
&s_i=2\cos\left(\frac{\pi}{N+1}i\right), \quad i=1,\dots,N, \label{eq:eig_A}\\
&V_j^{(i)}=\sqrt{\frac{2}{N+1}}\sin\left(\frac{\pi}{N+1}ij\right),
\quad i, j=1,\dots,N. \label{eq:eigV_A}
\end{align}
Recall that $H=Ve^{-St_{\text{f}}}(\int_{0}^{t_{\text{f}}}e^{St}V^TBB^TVe^{St}\dif t)e^{-St_{\text{f}}}V^T$.
Substituting this in Eqs.~(\ref{eq:eig_A}) and (\ref{eq:eigV_A}), after some
algebraic manipulation, we obtain
\begin{equation} \label{eq:H}
H=\frac{1}{N+1}VWPWV^T,
\end{equation}
where
\[W=\begin{bmatrix}
 \sin(\theta) & & & \\
 & \sin(2\theta) & &\\
 & & \ddots & \\
 & & & \sin(N\theta)
\end{bmatrix}_{N\times N}\]
and
\[
P_{jk}=\int_0^{2t_{\text{f}}}\,e^{-[\cos(j\theta)+\cos(k\theta)]t} \dif t
\]
with $\theta=\pi/(N+1)$, $j, k=1, \cdots, N$.

As a result, we have
$S=\mbox{diag}\{2cos(\frac{\pi}{N+1}),
2cos(\frac{2\pi}{N+1}),...,2cos(\frac{N\pi}{N+1})\}$.
The minimum eigenvalue of $H$ is given by
\begin{equation} \label{eq:eigH_inv}
\lambda_{H_N}=e^{-2cos(\frac{N\pi}{N+1})t_{\text{f}}}\lambda_{W_N}
e^{-2cos(\frac{N\pi}{N+1})t_{\text{f}}}=e^{-4cos(\frac{N\pi}{N+1})t_{\text{f}}}\lambda_{W_N}
=1/\lambda_1.
\end{equation}

The Rayleigh-Ritz theorem can be used to bound $P$ as:
\begin{equation}
\lambda_{P_N}\leq \frac{y^{T}Py}{y^{T}y}\leq \lambda_{P_1},
\end{equation}
where $y=[y_1,y_2,\dots,y_N]^T$ is an arbitrary nonzero column vector,
$\lambda_{P_N}$ and $\lambda_{P_1}$ are the maximal and minimal
eigenvalues of $P$, respectively. Letting $T=2t_{\text{f}}$, we have
\begin{align}
y^TPy &= (y_1 \cdots y_N) \left[\int_0^T\,e^{-[\cos(j\theta)+\cos(k\theta)]\tau}
\dif \tau \right]_{N\times N}\begin{pmatrix} y_1\\
\vdots\\
y_N
\end{pmatrix}\notag\\
&= \sum_{j, k=1}^N \, y_j y_k \int_0^T\,e^{-[\cos(j\theta)+\cos(k\theta)]\tau}
\dif \tau\notag\\
&= \sum_{j, k=1}^N \, \langle y_j e^{-[\cos(j\theta)t]},
y_k e^{-[\cos(k\theta)t]} \rangle \notag\\
&= \langle \sum_{j=1}^N\, y_j e^{-[\cos(j\theta)]t}, \sum_{j=1}^N\, y_j
e^{-[\cos(j\theta)]t}\rangle, \label{eq:RR_P}
\end{align}
with $\langle f,g\rangle \equiv \int_0^T\,f g \dif \tau$.

Letting $b_j = e^{-\cos(j\theta)}$ and performing a Taylor expansion of $b_j^t$
around $t=0$, we obtain
\begin{equation} \label{eq:Taylorbj}
b_j^t = \sum_{k=0}^{N-1}\,[-\cos(j\theta)]^k\frac{t^k}{k!}
+ [-\cos(j\theta)]^N\frac{t_j^N}{N!}
\end{equation}
with $t_j \in [0, T]$. Now letting
\begin{displaymath}
q_j(t) = \sum_{k=0}^{N-1}\,[-\cos(j\theta)]^k\frac{t^k}{k!},
\end{displaymath}
we have $b_j^t = q_j(t) + [-\cos(j\theta)]^N \cdot (t_j^N/N!)$.
Consequently, the numerator in the Rayleigh quotient can be expressed as
\begin{align}
y^TPy =& \langle \sum_{j=1}^N\,y_jb_j^t, \sum_{j=1}^N\,y_j b_j^t \rangle \notag\\
=& \langle \sum_{j=1}^N\,\left(y_jq_j(t) + y_j
[-\cos(j\theta)]^N\frac{t_j^N}{N!}\right), \sum_{j=1}^N\,\left(y_jq_j(t)
+ y_j [-\cos(j\theta)]^N\frac{t_j^N}{N!}\right)\rangle \notag\\
=& \langle \sum_{j=1}^N\,y_j q_j(t), \,\sum_{j=1}^N\,y_j q_j(t) \rangle
+ 2 \langle \sum_{j=1}^N\,\,y_j q_j(t), \sum_{j=1}^N\,y_j
[-\cos(j\theta)]^N\frac{t_j^N}{N!} \rangle \notag\\
&+ \langle \sum_{j=1}^N\,y_j [-\cos(j\theta)]^N\frac{t_j^N}{N!},\,
\sum_{j=1}^N\,y_j [-\cos(j\theta)]^N\frac{t_j^N}{N!} \rangle \notag\\
\leq & \underbrace{\langle \sum_{j=1}^N\,\,y_j q_j(t),
\sum_{j=1}^N\,y_j q_j(t) \rangle}_{\text{Denote as }K_1}
+ \underbrace{2\frac{T^N}{N!} \sum_{k=1}^N\,|y_k| \left(\left|
\langle \sum_{j=1}^N\,y_j q_j(t), 1 \rangle \right|\right)}_{\text{Denote as }K_2}
\notag\\
& + \underbrace{\left(\frac{T^N}{N!}\right)^2\sum_{j, k=1}^N\,
|y_j y_k|T}_{\text{Denote as}K_3}. \label{eq:approx_P}
\end{align}
Since $y=[y_1,y_2,\dots,y_N]^T$ is an arbitrary nonzero column vector, for each
$N$ and $T$, we can choose $y=y_m$ insofar as $K_1$ and $K_2$ are relatively
small compared with $K_3$. We can normalize $y_m^T y$ to arrive at
\begin{equation} \label{eq:eigH_N_T}
\lambda_{P_N} \leq \frac{y_m^TPy_m}{y_m^Ty_m} =
\left(\frac{T^N}{N!}\right)^2\sum_{j, k=1}^N\,
|y_{m_j} y_{m_k}|T\thicksim O\left(\frac{T^{2N}}{(N!)^2}\right),
\end{equation}
where $\lambda_{P_N}$ is the smallest eigenvalue of $P$. Recall that $P$
is symmetric and positive definite, using Cholesky decompostion we can
obtain its factorization~\cite{Strang:book} as $P=LL^T$, where $L$ is the
lower triangular matrix with its diagonal filled with square roots of
eigenvalues of $P$. Therefore, Eq.~(\ref{eq:H}) can be written as
$H=\frac{1}{N+1}V W LL^T W V^T$. Since orthonormal transform does not
change the eigenvalues of a matrix, $H$ has the same eigenvalues as
$R=\frac{1}{N+1}WLL^TW=\frac{1}{N+1}WL (WL)^T$. Suppose
$\Lambda_P=\mbox{diag}\{\lambda_{P_1},\lambda_{P_2},\dots,\lambda_{P_N}\}$ is the
diagonal eigenvalue matrix of $P$ in descending order. The $j$th eigenvalue
of $R$ satisfies
\begin{displaymath}
\lambda_{R_j}=\frac{1}{N+1}\lambda_{P_j} (\sin k\theta)^2\leq
\frac{1}{N+1}\lambda_{P_j},
\end{displaymath}
where $j$ and $k$ run from $1$ to $n$. The control energy $E(t_{\text{f}})$ can
then be approximated as
\begin{equation} \label{eq:chain_E_final}
E(t_{\text{f}}) \approx \lambda_{H_N}^{-1}\thicksim
O\left((N+1)\frac{(N!)^2}{t_{\text{f}}^{2N}}\right).
\end{equation}

\section*{Appendix D: Correlation between network control
energy and smallest eigenvalue of $H$-matrix}

\begin{figure}
\begin{center}
\epsfig{figure=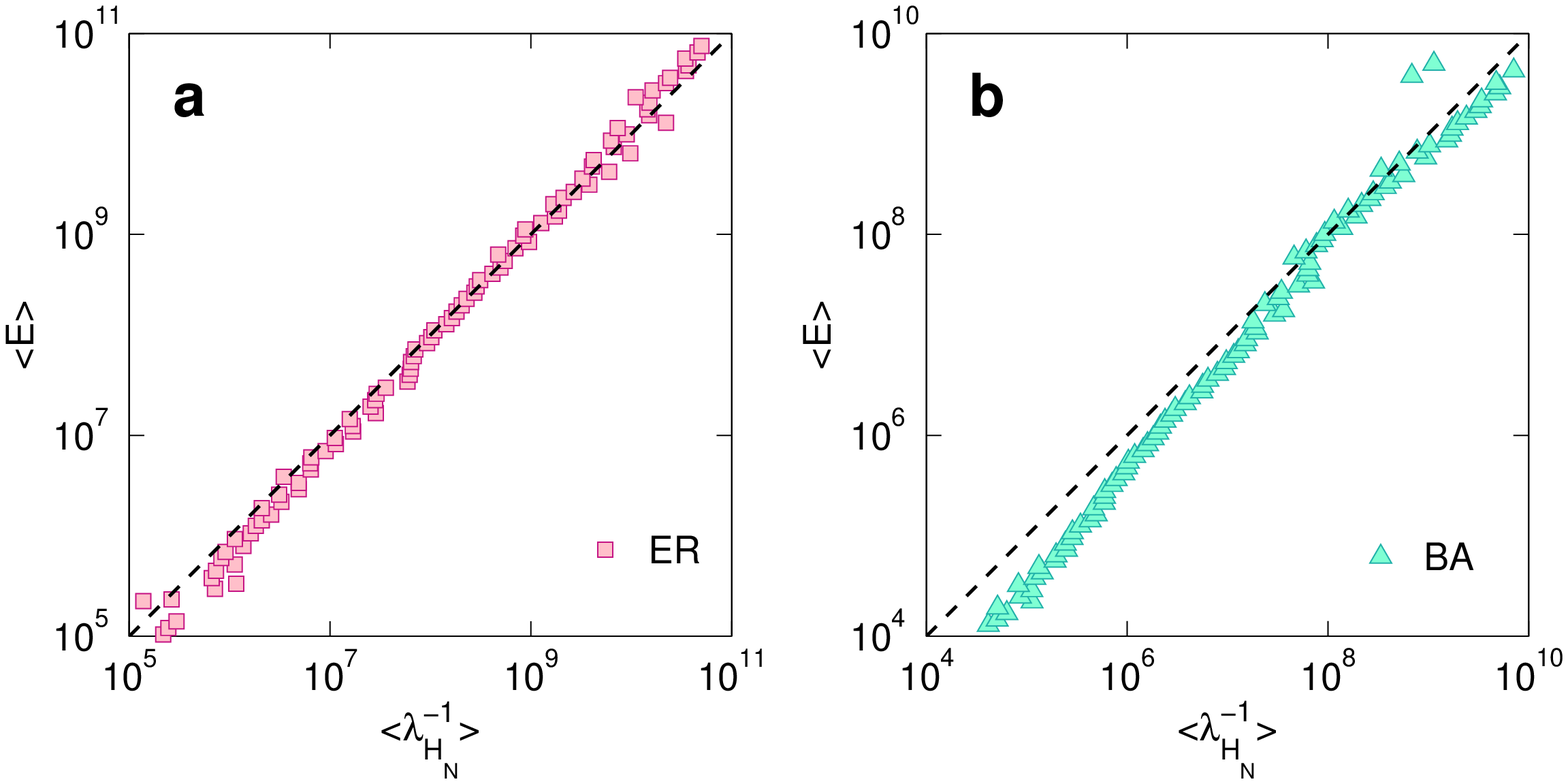,width=\linewidth}
\caption{\textbf{Correlation between network control energy and the smallest
eigenvalue of $H$-matrix}. Network size is $N=100$, directional link probability
between any pair of nodes is $P_{\text{b}}=0.1$, and average degree is (a) $\langle k\rangle=6$
for ER random networks and (b) $\langle k \rangle = 8$ for BA sale-free networks.}
\label{figS:Heign}
\end{center}
\end{figure}

Strong correlation between the average network control energy,
$\langle E\rangle$, and the smallest eigenvalue of the $H$-matrix,
$\lambda_{H_N}^{-1}$, for ER random and BA scale-free networks can
be observed in Fig.~\ref{figS:Heign}, indicating that the network
control energy is essentially determined by the smallest eigenvalue
of its $H$-matrix.

\newpage

\section*{Appendix E: LCC-skeleton and double-chain interaction models}

\paragraph*{E1: LCC-skeleton model.}
Referring to Fig.~4 in the main text, we assume that the control
chains are independent of each other so that $E_2$ is negligible as
compared with $E_1$. Each control chain is effectively a 1D string.
Due to the exponential increase in energy via chain
length increment, $E_1$ can be regarded as the sum of control energies
associated with the set of unidirectional 1D strings, to which the
contribution of the LCC dominates. The required energy to control the
full network can thus be approximated as that required to control all
LCCs,
\begin{equation} \label{E1_E2}
E = E_1 + E_2 \approx E_1 \approx m \cdot E_{L}
\approx m \cdot \lambda_{H_{L}}^{-1},
\end{equation}
where $E_L$ denotes the energy required to control an LCC,
$\lambda_{H_{L}}$ is the smallest eigenvalue of the LCC's $H$ matrix
$H_{L}$, and $m$ denotes the degeneracy (multiplicity) of the LCC, as
shown in Fig.~4(b) in the main text. Results presented in Fig.~3(b)
of the main text demonstrate a positive correlation between $E$ and
$m \cdot E_{L}$, reinforcing the idea the independent LCCs are the
key topological structure dictating the energy required to control the
whole network. In particular, if a network contains long LCCs (as can
be determined straightforwardly by maximum matching from the structural
controllability theory~\cite{LSB:2011}), there is high likelihood that
it cannot be practically controlled as practically the required energy
would diverge.

Reasoning from an alternative standpoint, an arbitrary combination of
$D_{\text{C}}$ and $m$ effectively represents a network, as shown in
Fig.~4(b) in the main text, and the entire network ensemble can be
represented by the ensemble of all possible combinations of LCCs. In
the LCC ensemble, the quantities $D_{\text{C}}$ and $m$ emerge according
to their probability density functions, $P_{D_{\text{C}}}(D_{\text{C}})$
and $P_m(m)$, respectively, and the appearance of an arbitrary pair of
$D_{\text{C}}$ and $m$ is determined by their joint probability density
function $P(D_{\text{C}}, m)$. Consequently, the distribution of the
energy required to control the original network can be characterized
accurately by the distribution of the energy required to control the
LCC skeleton in the corresponding ensemble.

\begin{figure}
\begin{center}
\epsfig{figure=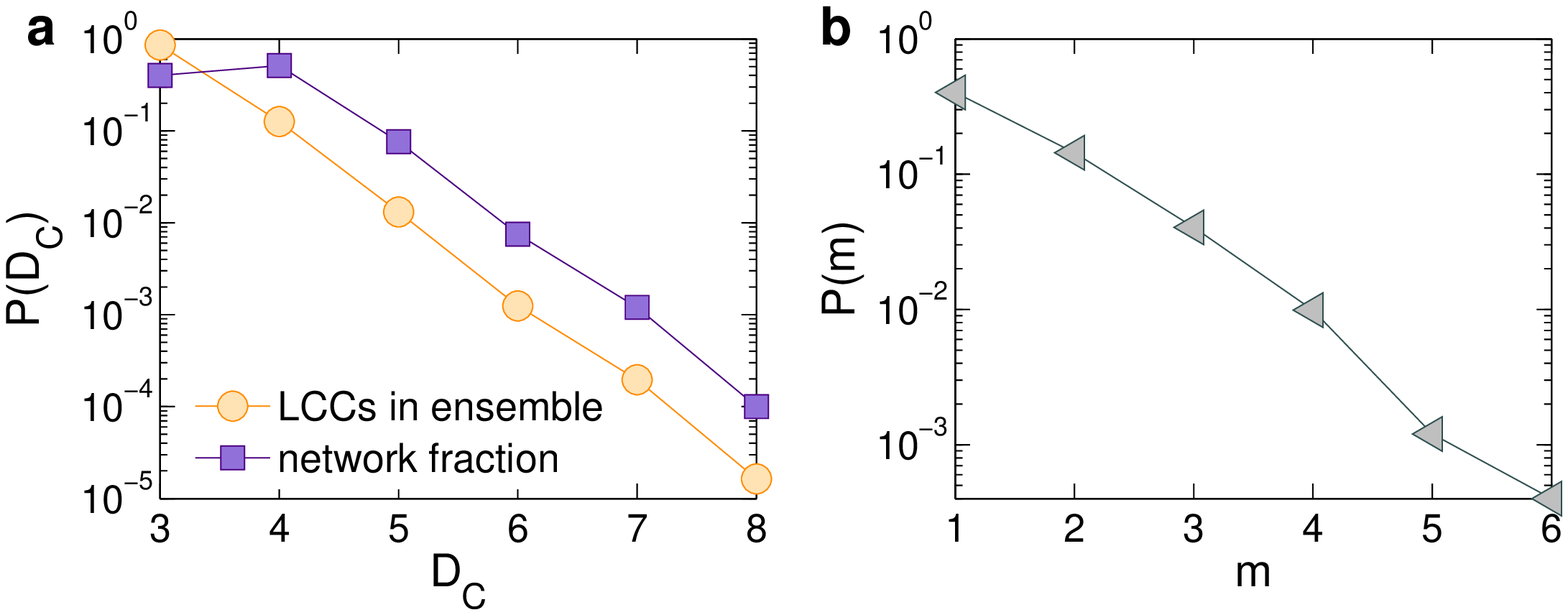,width=\linewidth}
\caption{\textbf{Probability distribution of control diameter and its degeneracy}.
(a) Distribution of $D_{\text{C}}$ (blue) and the probability that an LCC of
length $D_{\text{C}}$ appears in the random network ensemble versus $D_{\text{C}}$.
(b) Probability distribution of LCC-degeneracy $m$ for networks with
$D_{\text{C}} > 2$.}
\label{fig:LCC_distri}
\end{center}
\end{figure}

Figure~\ref{fig:LCC_distri}(a) shows the distribution of the control
diameter $D_{\text{C}}$, essentially the length distribution of LCCs.
The probability density function decays approximately  exponentially
with $D_{\text{C}}$, so we write
\begin{equation} \label{eq:P_DC}
P_{D_{\text{C}}}(D_{\text{C}}) = a \cdot e^{-b\cdot D_{\text{C}}},
\end{equation}
where $a$ and $b$ are positive constants. Using the relationship between
$E_{L}$ and $D_{\text{C}}$ [e.g., Fig.~3(a) in the main text], we have
\begin{equation} \label{eq:E_L}
E_{L} \approx A \cdot e^{B\cdot D_{\text{C}}} \Rightarrow D_{\text{C}}
\approx \frac{1}{B} \ln{\frac{E_{L}}{A}},
\end{equation}
where $A$ and $B$ are positive constants. The probability density function
of $E_L$ can then obtained as
\begin{equation} \label{eq:P_E_L}
P_L(E_{L}) = P_{D_{\text{C}}}( \frac{1}{B} \ln \frac{E_{L}}{A} )
\cdot |\frac{\dif (\frac{1}{B} \ln \frac{E_{L}}{A})}{\dif E_L}|
 \approx \frac{a}{B} A^{\frac{b}{B}} \cdot E_{L}^{-( 1 + \frac{b}{B})}.
\end{equation}
In the ER random network ensemble, the probability density of LCC
degeneracy $m$ for networks with $D_{\text{C}}>2$ also exhibits an
exponential decay, as shown in Fig.~\ref{fig:LCC_distri}(b):
\begin{equation} \label{eq:P_m}
P_m(m) = c \cdot e^{-g \cdot m},
\end{equation}
where $c$ and $d$ are positive constants.

Since the control energy depends monotonously on the control diameter
$D_{\text{C}}$, the energy dependence on $m$ can be revealed by
examining the correlation between $D_{\text{C}}$ and $m$, which can
in general be either positive or negative. From Eq.~(\ref{eq:P_m}), we
see that $P_m(m)$ increases exponentially with $m$, implying a positive
correlation:
\begin{equation}
D_{\text{C}}  = s_1 + s_2\cdot m,
\end{equation}
where $s_1$ and $s_2$ are positive constants. This form of relation ensures
that $P_m(m)$ has the form in Eq.~(\ref{eq:P_m}). Positive correlation, however,
means that the number of LCCs increases with its length, which is unphysical
for random networks. These arguments suggest that a contradiction can arise
if we assume either positive or negative correlation between $D_{\text{C}}$
and $m$. A natural resolution is that these two quantities are independent
of each other.
Since $E_L$, the energy required
to control a chain of length $L$, is determined mainly by the control
diameter $D_{\text{C}}$, $E_L$ and $m$ can be assumed to be independent of
each other so that their joint probability density function can be expressed as
$P(E_L, m) \approx P_L(E_L)\cdot P_m(m)$.

Having obtained $P_L(E_L)$ and $P_m(m)$, we can calculate the cumulative
probability distribution function of the estimated control energy
$E = m \cdot E_L$ required to control the original network. We have
\begin{eqnarray}\label{eq:F_E}
&& F_E(E)  = P(m \cdot E_L < E)
= \int_0^{\infty} [\int_0^{\frac{E}{E_L}} P(E_L, m) \cdot \dif m] \dif E_L \\ \nonumber
&&= \int_0^{\infty} [\int_0^{\frac{E}{E_L}} P_L(E_L)
\cdot P_m(m) \cdot \dif m] \dif E_L
\approx \frac{caA^{\frac{b}{B}}}{gB} \cdot \{-\frac{b}{B} - [\Gamma (\frac{b}{B}) -
\Gamma (\frac{b}{B}, gE)] \cdot (gE)^{-\frac{b}{B}}  \},
\end{eqnarray}
where $\Gamma (\frac{b}{B})$ and $\Gamma (\frac{b}{B}, gE)$ are the Gamma and
incomplete Gamma function, respectively. Thus, the probability density function
of $E$ can then be expressed as
\begin{equation} \label{eq:P_E}
P_E(E) = \frac{\dif F_E(E)}{\dif E}
\approx \frac{caA^{\frac{b}{B}}}{gB} \cdot \{-\frac{e^{-gE}}{E}
+ [\Gamma (\frac{b}{B}) - \Gamma (\frac{b}{B},
gE)] \cdot (gE)^{-(1+\frac{b}{B})}  \},
\end{equation}
where the first term $-e^{-gE}/E$ can be neglected due to the typically
large value of $E$. Since we observe numerically that the difference between the
two Gamma functions is approximately constant:
$h(\Gamma) \equiv \Gamma (\frac{b}{B}) - \Gamma (\frac{b}{B}, gE) \approx 1.7$,
we can simplify Eq.~(\ref{eq:P_E}) as
\begin{equation}\label{eq:P_E_final}
P_E(E) \approx C \cdot E^{-(1+\frac{b}{B})},
\end{equation}
where $C = [\frac{caA^{\frac{b}{B}}}{gB} \cdot g^{-(2+\frac{b}{B})} \cdot h_{\Gamma}]$
is a positive constant. Equation~(\ref{eq:P_E_final}) indicates a power-law
distribution of the control energy, providing an analytical explanation to the
numerically discovered energy distribution for practically controllable networks,
as exemplified in Fig.~3 in the main text. To get a rough idea about the value
of the power-law scaling exponent, say we take $B \approx 2$ and $b\approx 1$
(typical numerical values). A theoretical estimate of the power-law exponent is thus
$1+b/B \approx 1.5$, which is consistent with the value obtained from
results from direct numerical simulation. The fact that the distribution of
$E_L$ is power law with the identical exponent provides additional support for
our assumption that the LCC degeneracy $m$ plays little role in determining the
control energy. It is the combination of the exponential decay in the probability
distribution of the control diameter [cf., Eq.~(\ref{eq:P_DC})] and the exponential
increase in the energy required to control LCC with its length [cf., Eq.~(\ref{eq:E_L})]
that gives rise to the power-law energy distribution of the LCCs, which ultimately
leads to the power-law distribution in the actual energy required to control the
original random network.

We see that the control diameter of a network is a key quantity determining
the required control energy. The topological diameter, on the other hand,
is a fundamental quantity characterizing, for example, the small-world
structure of the network~\cite{WS:1998}. An interesting issue concerns
the relation between the control and topological diameters. In particular,
if the network has a large diameter, does it mean that its control diameter must
be large as well? This issue has been addressed, with the finding that there is
little correlation between the two types of diameters.

\paragraph*{E2: Double-chain interaction model.}
Our analysis of the LCC-skeleton model predicts power-law distribution
of the required energy for practically controllable networks, which agrees
qualitatively with numerics. However, in the model interactions among the
coexisting chains are ignored. In a physical system, interactions among
the basic components usually plays an important role in determining
the system's properties. To obtain a more accurate estimate of the behaviors
of the control energy, we need to include the interactions among the
chains. The necessity is further justified as there are discrepancies
between the actual control energy and that from the LCC-skeleton model,
as exemplified in Fig.~3(b) in the main text. In particular,
there is an approximately continuous distribution in the energy required
to control the actual network, but the distribution of the energy from
the LCC-skeleton model tends to aggregate into a number of subintervals,
each corresponding to a certain value of the control diameter associated
with an LCC. Thus, in order to reproduce the numerically obtained energy
distributions, we must incorporate the interactions among the LCCs into the
model. However, including the interactions makes analysis difficult, as
there are typically a large number of interacting pairs of chains. To gain
insight into the role played by the interactions, it is useful to focus on
the relatively simple case of two interacting chains.

\begin{figure}
\begin{center}
\epsfig{figure=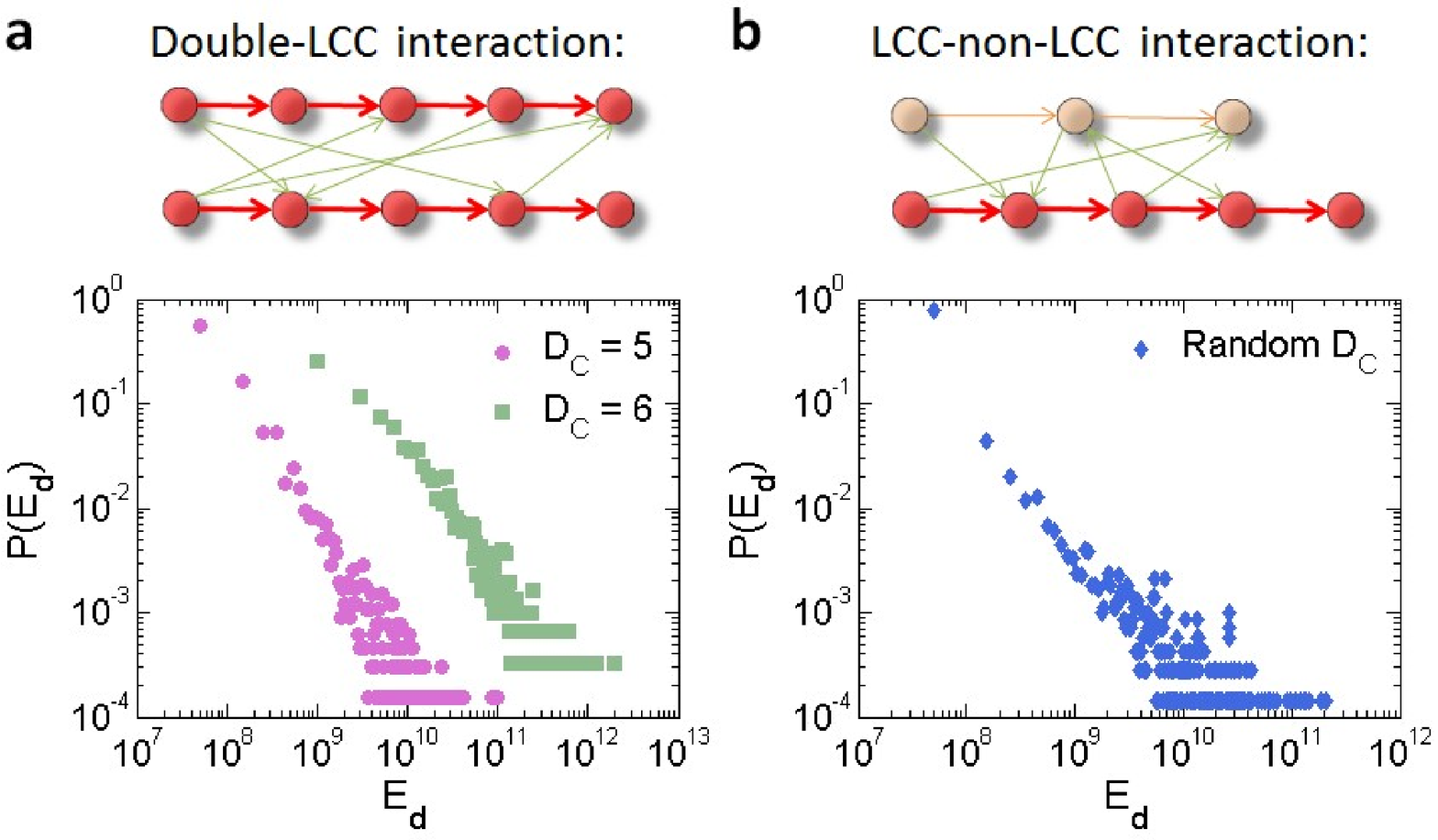,width=\linewidth}
\caption{ \textbf{Distribution of control energy in a double-chain
interaction models.} (a) Two chains of $D_{\text{C}} = 5$ (red) and
$D_{\text{C}} = 6$ (black), where the upper panel is a schematic
illustration of two LCCs of identical length $D_{\text{C}} = 5$
interacting with each other via some random links (green) between
them. (b) Two chains with their lengths randomly chosen from $3-6$,
where the upper panel shows the case of two interacting chains of
length $5$ and $3$. The longer chain (red) plays the role of LCC,
while the shorter chain is a non-LCC (orange).}
\label{fig:double_chain}
\end{center}
\end{figure}

Our double-chain interaction model is constructed, as follows. Consider two
identical unidirectional chains, denoted by $C1$ and $C2$, each of length
$D_{\text{C}}$. Every node in $C1$ connects with every node in $C2$ with
probability $p$, all links between the two chains are unidirectional. A
link points to $C2$ from $C1$ with probability $p_{1\rightarrow2}$ and the
probability for a link in the opposite direction is
$p_{2\rightarrow1} = 1-p_{1\rightarrow2}$.
By changing the connection rate $p$ and the directional bias
$p_{1\rightarrow2}$, we can simulate and characterize various interaction
patterns between the two chains. To be concrete, we generate an ensemble
of 10000 interacting double-chain networks, each with $2\cdot D_{\text{C}}$ nodes
and multiple randomized interchain links as determined by the parameters
$p$ and $p_{1\rightarrow2}$. As shown in Fig.~\ref{fig:double_chain}(a), the
distribution of the control energy displays a remarkable similarity
to that for random networks, in that a power-law scaling behavior emerges
with the exponent about $1.5$. A striking result is that the energy distributions
from the double-chain interaction model are much more smooth than those from
the LCC-skeleton model, indicating the key role played by the interchain
interactions in spreading out the control energies that are clustered when
the interactions are absent. The power-law distribution holds robustly
with respect to variations in the parameters $p$ and $p_{1\rightarrow2}$.
In addition, to reveal the role of the interaction between an LCC and a
non-LCC chain in the control energy, we randomly pick their lengths
from $[3,6]$ with equal probability, where the longer chain acts as an LCC.
Again, we observe a strong similarity between the energy distributions
from random networks and from this model, as shown in Fig.~\ref{fig:double_chain}(b),
suggesting a universal pattern followed by pair interactions, regardless of
the length of the chains. In particular, interactions between two chains,
LCC or not, have similar effect on the control-energy distribution.
These results indicate that the double-chain interaction model captures
the essential physical ingredients of the energy distribution in controlling
complex networks.

\section*{Appendix F: The practical controllability of real-world networks}
Table~\ref{table:real_world_data} lists the names and types of the real-world
networks studied and Table~\ref{table:real_world} presents more detailed 
information about the controllability of the $18$ real-world networks 
analyzed in the main text.

\begin{table*}
\caption{Description of the 18 real-world networks used in the
paper ($N$ - number of nodes; $M$ - number of edges).}
\begin{center}
\setlength{\extrarowheight}{3pt}
\noindent\makebox[\textwidth]
{%
\centering
  \begin{tabularx}{\linewidth}{c c c c c c}

    \cline{1-6}

  Type& Index & Name & $N$ & $M$ & Description \\ \cline{1-6}
  \multirow{2}{*}{\begin{minipage}{0.6in}\centering Trust\end{minipage}}
  &1&College Student \cite{Van:2003,Milo:2004} & 32 & 96 & Social network \\
  &2&Prison Inmate \cite{Van:2003,Milo:2004}& 67 & 182 & Social network \\
  \\
  \multirow{3}{*}{\begin{minipage}{1.6in}\centering Circuits\end{minipage}}
  &3  & s208a \cite{Milo:2002} &122& 189 & Logic circuit \\
  &4  & s420a \cite{Milo:2002} &252& 399 & Logic circuit \\
  &5  & s838a \cite{Milo:2002} &512& 189 & Logic circuit \\
  \\
  \multirow{2}{*}{\begin{minipage}{1.6in}\centering Citation\end{minipage}}
  &6  &Small World \cite{WS:1998}&233& 1988 & Stanley Milgram \\
  &7  & Kohonen \cite{Davis:2011}&3772& 96 & T. Kohonen \\
  \\
  \multirow{3}{*}{\begin{minipage}{0.6in}\centering Protein\end{minipage}}
  &8  &Protein-1 \cite{Milo:2004} &95& 213 & Protein network \\
  &9  &Protein-2 \cite{Milo:2004} &53& 123 & Protein network \\
  &10  &Protein-3 \cite{Milo:2004} &99& 212 & Protein network \\
  \\
  \multirow{7}{*}{\begin{minipage}{1.6in}\centering Food Web\end{minipage}}
  &11 & St. Martin \cite{Baird:1998} & 45 & 224 & Food Web \\
  &12 & Seagrass \cite{Christ:1999} & 49 & 226 & Food Web  \\
  &13 &Grassland \cite{Dunne:2002} & 88 & 137 & Food Web \\
  &14 &Ythan \cite{Dunne:2002}      &135& 601 & Food Web \\
  &15 &Silwood    \cite{Memmott:2000} &154& 370 & Food Web \\
  &16 &Little Rock \cite{Martinez:1991} &183& 2494 & Food Web \\
  &17 &Baydry \cite{Ulan:2005} &128& 2137 &Food Web \\
  &18 &Florida \cite{Ulan:2005} &128& 2106 &Food Web \\
  \cline{1-6}
  \end{tabularx} }
  \label{table:real_world_data}
\end{center}
\end{table*}

\begin{table*}[ht]
\caption{Practical controllability of real-world networks studied in
Ref.~\cite{LSB:2011}, where $N_\text{D}$ denotes the number of controllers as
determined by the structural controllability theory and $M^\star$ is
the number of augmented driver nodes needed to make the network
practically controllable. The densities of the original driver nodes
and with $M^\star$ augmented drivers included are
$n_{\text{D}}=N_\text{D}/N$ and
$n_{\text{D}}^\star=N_\text{D}^\star\left. \right/ N$,
respectively, where $n_{\text{D}}^\star$
is the new measure of controllability, and $E^\star$
denotes the new energy. When an additional control signal is added to
the middle of each LCC [strategy (I)] so that $M^{\star}_{\text{mid}}$
extra control inputs are used, the control energy is $E^{\star}_{\text{mid}}$.
We also test another strategy in which an extra control signal is
applied to each of the $m$ convergent nodes of all LCCs [strategy (II)],
in which $M^{\star}_{\text{end}}$ additional control inputs are used. The
control energy required is denoted as $E^{\star}_{\text{end}}$. The control diameter $D_{\text{C}}$ of each network is listed in the last column.
(See Table A1 in Appendix F for detailed description of the 18 networks.)}
  \begin{tabular*} {\hsize}
  {@{\extracolsep{\fill}}rrrrrrrrrrrrr}
  \cline{1-13}
  \cr
  \multicolumn1c{Type} & \multicolumn1c{Name} & \multicolumn1c{$N$} & \multicolumn1c{$N_\text{D}$} & \multicolumn1c{$M^\star$}  & \multicolumn1c{$n_{\text{D}}$} & \multicolumn1c{$n_{\text{D}}^\star$} & \multicolumn1c{$E^\star$} & \multicolumn1c{$M^{\star}_{\text{mid}}$} & \multicolumn1c{$E^{\star}_{\text{mid}}$} & \multicolumn1c{$M^{\star}_{\text{end}}$} & \multicolumn1c{$E^{\star}_{\text{end}}$} & \multicolumn1r{$D_{\text{C}}$}\\
  \cr
  \cline{1-13}
  \cr
  \multicolumn1c{Trust}
  & Coll. Student & 32 & 6 & 0 & 0.19 & 0.19 & $7.9\times10^{10}$ & 3 & $2.5\times10^{5}$ & 4 & $7.3\times10^4$ & 4\cr
  & Prison Inmate & 67 & 9 & 27 & 0.13 & 0.54 & $2.3\times 10^{7}$ & 1 & $4.6\times10^3$ & 1 & $9.7\times 10^4$ &5\cr
  \cr
   \multicolumn1c{Electronic Circuits}
    & s208a &122& 29 & 40 & 0.24 & 0.57 & $2.3\times10^7$ & 1  &$3.0\times10^5$  & 1 & $4.0\times10^5$ & 5\cr
    & s420a &252& 59 & 71 & 0.23 & 0.52 & $2.6\times10^6$ &11 &$2.1\times10^5$  & 10& $2.2\times10^5$  & 4\cr
    & s838a &512& 119 &81& 0.23 & 0.39 & $4.1\times10^9$ &13 &$3.1\times10^6$  & 6  & $3.7\times10^6$ & 5\cr
  \cr
  \multicolumn1c{Citation}
    &Small World &233& 140 &11& 0.60 & 0.65 & $2.0\times10^3$ &1 &$1.9\times10^3$ & 1  & $1.9\times10^3$ & 5\cr
    & Kohonen &3772& 2114 &413& 0.56 & 0.67 & $9.4\times10^4$ &49 &$4.0\times10^4$ & 37 & $5.1\times10^4$ & 3\cr
    \cr
  \multicolumn1c{Protein}
    &Protein-1 &95& 48 & 19 & 0.51 & 0.71 & $1.9\times10^{10}$ &7 &$8.5\times10^2$ & 5  & $2.2\times10^3$  & 3\cr
    &Protein-2 &53& 13 & 0   & 0.25 & 0.25 & $3.7\times10^9$ & 2 &$9.0\times10^8$ & 2 & $1.2\times10^9$ & 4\cr
    &Protein-3 &99& 22 & 35 & 0.22 & 0.58 & $3.9\times10^5$ & 3 &$5.1\times10^4$ & 3 & $5.3\times10^4$ &4\cr
    \cr
  \multicolumn1c{Food Web}
   & St. Martin  & 45 & 14 & 9 & 0.31 & 0.51 & $2.7\times10^3$ & 2 &$9.2\times10^2$ & 2 & $1.5\times 10^3$ & 3\cr
   & Seagrass & 49 & 13 & 7 & 0.27 & 0.41 & $3.6\times10^3$ & 3 &$8.3\times10^2$ & 2 & $3.4\times 10^3$ & 3\cr
   &Grassland & 88 & 46 & 0 & 0.52 & 0.52 & $3.4\times10^5$ & 1 &$8.6\times10^4$ & 1 & $9.2\times 10^4$ & 4\cr
   &Ythan      &135& 69 & 14 & 0.51 & 0.62 & $2.6\times10^3$ & 4 &$1.9\times10^3$ & 2 & $2.0\times 10^3$ & 3\cr
   &Silwood    &154& 116 & 12 & 0.75 & 0.83 & $1.5\times10^4$ & 3 &$1.0\times10^3$ & 2 & $9.9\times 10^2$ & 3\cr
   &Little Rock &183& 99 & 36 & 0.54 & 0.74 & $5.4\times10^3$ & 48 &$3.6\times10^3$ & 48 & $3.6\times10^3$ & 2\cr
   &Baydry &128& 62 & 34 & 0.48 & 0.75 & $1.7\times10^3$ & 32 &$9.6\times10^2$ & 48 & $9.6\times10^2$ & 2\cr
   &Florida &128& 30 & 69 & 0.23 & 0.77 & NaN & 1 &NaN & 1 & NaN & 3\cr
  \cr
  \cline{1-13}
  \end{tabular*}
  \label{table:real_world}
\end{table*}

\newpage

\begin{figure}
\begin{center}
\epsfig{figure=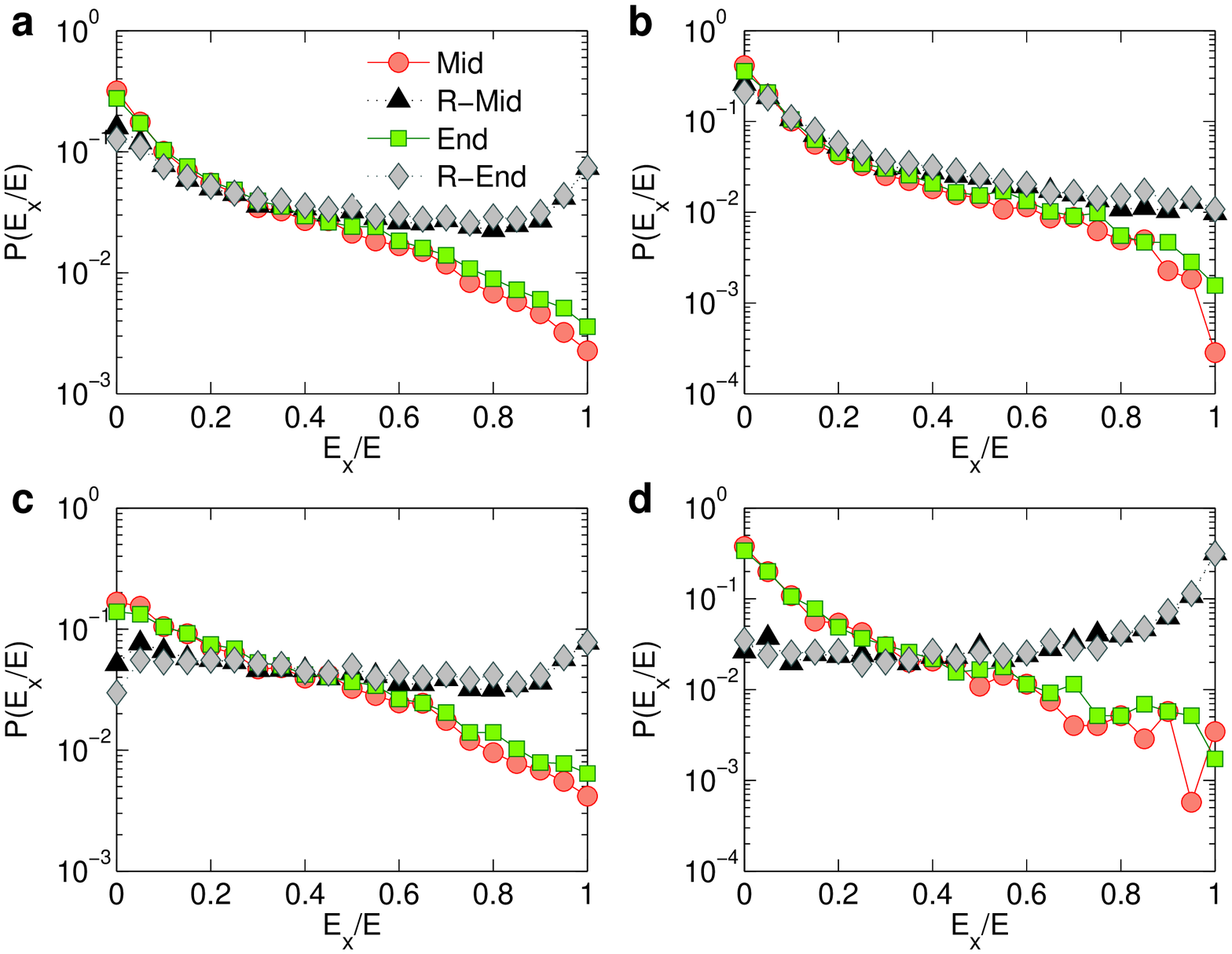,width=\linewidth}
\caption{\textbf{Effects of redundant control inputs.}
(a) For control diameter $D_{\text{C}} = 2$, distribution of the
energy ratio $E_x/E$ under optimization strategies
(I) (Mid, red circles) and (II) (End, green squares), where $E_x$ and $E$
are the required energies with and without redundant control, respectively.
Results from two randomized optimization strategies are marked by R-Mid
(black triangles) and R-End (gray diamonds), corresponding to strategies
(I) and (II), respectively. The values of $E_x/E$ are collected from
practically controllable networks from an ensemble of $10000$ ER random
networks ($\langle k \rangle = 6$, $P_{\text{b}} = 0.1$). For each network, if
strategy (I) [or (II)] requires $r$ redundant controls, $r$ additional
random control inputs are applied to the system $10$ times to average
out the random fluctuations. Panels (b-d) show the $E_x/E$ distributions for
networks with control diameter $D_{\text{C}} = 3$, $4$, or $5$, respectively.}
\label{fig:redundancy}
\end{center}
\end{figure}

\section*{Appendix G: Energy optimization of modeled complex networks and a cascade parallel R-C circuit
network}

\paragraph*{G1: Optimization strategy for modeled networks.}
A realistic complex network can often have multiple LCCs,
requiring multiple redundant control inputs.
Say we wish to introduce a small number of extra control signals. Due to
the $m$ degeneracy in the end nodes LCCs, it seems that the number of
redundant control inputs should exceed $m$ if every LCC receives one such signal.
However, since even a unity deduction in the LCC length
can significantly lower the control energy, a simpler strategy is to place one
redundant control input at each of the $m$ end-nodes to which all possible LCCs
converge. In this case, each LCC in the network is broken into a chain of length
$L-1$ and a single node, and consequently, the control energy is now determined
by one-dimensional chains of length $L-1$ instead of length $L$.
Figure~\ref{fig:redundancy}(a) shows the effects of two optimization strategies
to introduce redundant control signals on the energy distribution: applying
one redundant control signal (I) at the middle and (II) at the end of each and
every LCC, respectively. For comparison, for each strategy, the same number
of redundant control inputs are also applied randomly throughout the network.
The ratio between the control energy under optimization strategy, $E_x$, and
the original control energy $E$ characterizes the effectiveness of the
optimization strategies. In particular, if the distribution of $E_x/E$
is concentrated on small values of $E_x/E$, then the corresponding optimization
strategy can be deemed to be effective. As shown in Fig.~\ref{fig:redundancy}(a),
both optimization strategies outperform the random strategies, with strategy (I)
performing slightly better than (II). The networks requiring proper optimization
to be practically controlled are typically those with long control diameters.
Figures~\ref{fig:redundancy}(b-d) show that this is indeed the case.

\begin{figure}
\begin{center}
\epsfig{figure=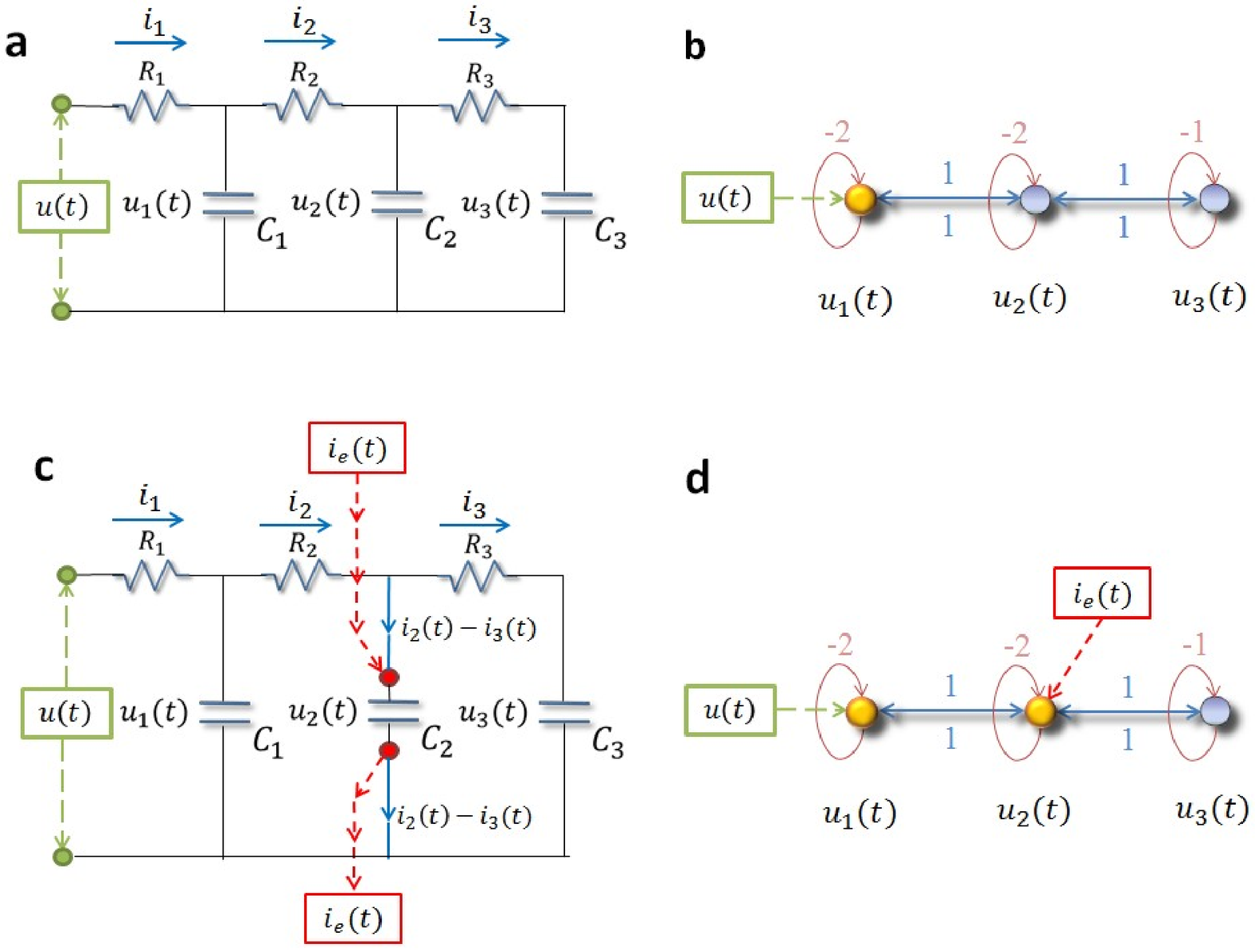,width=\linewidth}
\caption{\textbf{Controlling and optimizing a cascade parallel R-C circuit and the corresponding
network presentation.} (a) A cascade parallel R-C circuit with $3$
resistors ($R_1$, $R_2$, and $R_3$, each of resistance $1 \Omega$)
and $3$ capacitors ($C_1$, $C_2$, and $C_3$, each of capacitance $1$F),
where $u(t)$ is the external input voltage, $u_1(t)$, $u_2(t)$, and $u_3(t)$
are the voltages on the capacitors $C_1$, $C_2$, and $C_3$, respectively,
$i_1(t)$, $i_2(t)$, and $i_3(t)$ are the currents through the resistors
$R_1$, $R_2$, and $R_3$, respectively. (b) Network representation of
the circuit in (a). (c) Circuit with an extra external current input
$i_{\text{e}}(t)$ into the capacitor $C_2$. (d) The extra external current input
$i_{\text{e}}(t)$ serves as a redundant control input injected into node $2$ of
the network in (b). There are two driver nodes (yellow) in the network:
$1$ and $3$.}
\label{fig:sup_circuit}
\end{center}
\end{figure}

\paragraph*{G2: An example of controlling and optimizing a circuit system.}
We consider a cascade parallel R-C circuit consisting of three identical
resistors and capacitors as an example to illustrate how the circuit can
be abstracted into a directed network, as shown in Fig.~\ref{fig:sup_circuit}.
For convenience, we set $R_1 = R_2 = R_3 = R $ and $C_1 = C_2 = C_3 = C $,
and denote the currents through $R_1$, $R_2$, and $R_3$ as $i_1(t)$ , $i_2(t)$,
and $i_3(t)$, respectively. The equations of the circuit are
\begin{equation} \label{eq:circuit_1}
\left\{
\begin{array}{cl}
& u(t) = i_1(t) R + u_1(t) \\
& u_1(t) = i_2(t) R + u_2(t) \\
& u_2(t) = i_3(t) R + u_3(t) \\
& C \frac{\dif u_1(t)}{\dif t} = i_1(t) - i_2(t) \\
& C \frac{\dif u_2(t)}{\dif t} = i_2(t) - i_3(t) \\
& C \frac{\dif u_3(t)}{\dif t} = i_3(t)
\end{array}
\right.
\end{equation}
After some algebraic manipulation, we have
\begin{equation} \label{eq:circuit_2}
\left\{
\begin{array}{cl}
& \frac{\dif u_1(t)}{\dif t} = -\frac{2}{RC} u_1(t) + \frac{1}{RC}u_2(t) + \frac{1}{RC} u(t) \\
& \frac{\dif u_2(t)}{\dif t} = \frac{1}{RC} u_1(t) - \frac{2}{RC}u_2(t) + \frac{1}{RC} u_3(t) \\
& \frac{\dif u_3(t)}{\dif t} = \frac{1}{RC} u_2(t) - \frac{1}{RC}u_3(t),
\end{array}
\right.
\end{equation}
which can be written as
\begin{equation} \label{eq:circuit_3}
\left(
\begin{array}{c}
\frac{\dif u_1(t)}{\dif t} \\
\frac{\dif u_2(t)}{\dif t} \\
\frac{\dif u_3(t)}{\dif t} \\
\end{array}
\right) =
\left(
\begin{array}{ccc}
-\frac{2}{RC} & \frac{1}{RC} & 0 \\
\frac{1}{RC} & -\frac{2}{RC} & \frac{1}{RC} \\
0 & \frac{1}{RC} & -\frac{1}{RC} \\
\end{array}
\right)
\left(
\begin{array}{c}
u_1(t)\\
u_2(t) \\
u_3(t) \\
\end{array}
\right)
+
\left(
\begin{array}{c}
\frac{1}{RC}\\
0 \\
0 \\
\end{array}
\right) u(t).
\end{equation}
Setting $R = 1\Omega$ and $C = 1F$, we have
\begin{equation} \label{eq:circuit_4}
\left(
\begin{array}{c}
\frac{\dif u_1(t)}{\dif t} \\
\frac{\dif u_2(t)}{\dif t} \\
\frac{\dif u_3(t)}{\dif t} \\
\end{array}
\right) = A \cdot
\left(
\begin{array}{c}
u_1(t)\\
u_2(t) \\
u_3(t) \\
\end{array}
\right) +
B \cdot u(t),
\end{equation}
where
\begin{equation}
A = \left(
\begin{array}{ccc}
-2 & 1 & 0 \\
1 & -2 & 1 \\
0 & 1 & -1 \\
\end{array}
\right)
\end{equation}
is the adjacency matrix of the network representing the circuit, and
\begin{equation}
B = \left(
\begin{array}{c}
1\\
0 \\
0 \\
\end{array}
\right)
\end{equation}
is the control input matrix. The circuit has then been transferred
into a $3$-node bidirectional 1D chain network with adjacency matrix $A$.

Without loss of generality, we inject an extra external current input
$i_{\text{e}}(t)$ into the capacitor $C_2$, and the circuit equations become:
\begin{equation} \label{eq:circuit_5}
\left\{
\begin{array}{cl}
& u(t) = i_1(t) R + u_1(t) \\
& u_1(t) = i_2(t) R + u_2(t) \\
& u_2(t) = i_3(t) R + u_3(t) \\
& C \frac{\dif u_1(t)}{\dif t} = i_1(t) - i_2(t) \\
& C \frac{\dif u_2(t)}{\dif t} = i_2(t) - i_3(t) + i_{\text{e}}(t) \\
& C \frac{\dif u_3(t)}{\dif t} = i_3(t)
\end{array}
\right.
\end{equation}
The state equations are
\begin{equation} \label{circuit}
\left(
\begin{array}{c}
\frac{\dif u_1(t)}{\dif t} \\
\frac{\dif u_2(t)}{\dif t} \\
\frac{\dif u_3(t)}{\dif t} \\
\end{array}
\right) = A \cdot
\left(
\begin{array}{c}
u_1(t)\\
u_2(t) \\
u_3(t) \\
\end{array}
\right) +
B_{\text{e}}
\left(
\begin{array}{c}
u(t)\\
i_{\text{e}}(t) \\
\end{array}
\right) ,
\end{equation}
where
\begin{equation}
B_{\text{e}} = \left(
\begin{array}{cc}
1 & 0 \\
0 & 1 \\
0 & 0 \\
\end{array}
\right)
\end{equation}
is the control input matrix of the circuit under the original control
input $u(t)$ on node $1$ and a redundant control input $i_{\text{e}}(t)$
to node $2$. Similarly, the redundant control input can be
injected into any capacitor.

It is necessary to keep all other nodes unaffected while introducing
exactly one extra control input into the circuit. However, any
additional voltage change in any part of the circuit can lead
to voltage changes on all the capacitors. A change in the current
through a capacitor will not affect the currents in other
components of the network, since only the time derivative of its
voltage is affected. Thus, a meaningful way to introduce an
extra control signal input to one node of a circuit's network
is to inject current into one particular capacitor in the circuit.


\newpage

%
\end{document}